\documentclass[%
 reprint,
 amsmath,amssymb,
 aps,
]{revtex4-2}
\usepackage{lineno}
\usepackage{graphicx}
\usepackage{dcolumn}
\usepackage{bm}

\usepackage{xcolor}
\begin{document}


\title{Measurements of evaporation residue cross-sections and evaporation residue-gated $\gamma$-ray fold distributions for $^{32}$S+$^{154}$Sm system}
\author{R. Sariyal$^{1,2,^{*,\delta}}$, I. Mazumdar$^{2,^{\dagger}}$,  D. Mehta$^{1}$, N. Madhavan$^{3}$, S. Nath$^{3}$, J. Gehlot$^{3}$, Gonika$^{3}$, S. M. Patel$^{2}$, P. B. Chavan$^{2}$, S. Panwar$^{4}$, V. Ranga$^{4}$, A. Parihari$^{5}$, A. K. Nasirov$^{6,7}$, B. M. Kayumov$^{7,8}$ }
 \affiliation{ $^1$Department of Physics, Panjab University, Chandigarh-160014, India}
 \email{ranjansariyal17@gmail.com}


\affiliation{$^{2}$Department of Nuclear and Atomic Physics, Tata Institute of Fundamental Research, Colaba-400005, Mumbai, India}%
\email{indra@tifr.res.in\\$^{\delta}${Present Address: Department of Applied Sciences, Chandigarh Engineering College-CGC, Landran, Punjab-140507, India}}

\affiliation{$^{3}$Inter-University Accelerator Centre, Aruna Asaf Ali Marg, New Delhi-110067, India}
\affiliation{$^{4}$Department of Physics, Indian Institute of Technology, Roorkee, Roorkee-247667, Uttarakhand, India}
\affiliation{$^{5}$Department of Physics and Astrophysics, University of Delhi, New Delhi-110007, India}
\affiliation{$^{6}$BLTP, Joint Institute for Nuclear Research, Joliot-Curie 6, Dubna-141980, Russia}
\affiliation{$^{7}$Institute of Nuclear Physics, Uzbekistan Academy of Sciences, Tashkent-100214, Uzbekistan}%
\affiliation{$^{8}$New Uzbekistan University, Tashkent-100007, Uzbekistan}%
\date{\today}
\begin{abstract}

Evaporation Residue (ER) cross-sections and ER-gated $\gamma$-ray fold distributions are measured for the $^{32}$S + $^{154}$Sm nuclear reaction above the Coulomb barrier at six different beam energies from 148 to 191 MeV. $\gamma$-ray multiplicities and spin distributions are extracted from the ER-gated fold distributions. The ER cross-sections measured in the present work are found to be much higher than what was reported in a previous work using a very different target-projectile ($^{48}$Ti + $^{138}$Ba) combination, leading to the same compound nucleus $^{186}$Pt, with much less mass asymmetry in the entrance channel than the present reaction. This clearly demonstrates the effect of the entrance channel on ER production cross-section. The ER cross-sections measured in the present work are compared with the results of both the statistical model calculations and the dynamical model calculations. Statistical model calculations have been performed to generate a range of parameter space for both the barrier height and Kramers' viscosity parameter over which the ER cross-section data can be reproduced. The calculations performed using the dinuclear system (DNS) model reproduce the data considering both complete and incomplete fusion processes. DNS calculations indicate the need for the inclusion of incomplete fusion channel at higher energies to reproduce the ER cross-sections.
\end{abstract}

\maketitle

\section{Introduction}

Studies in heavy-ion-induced fusion-fission reactions continue to be at the forefront of research in low, and medium-energy nuclear physics. Understanding the complex dynamics involved in fusion-fission reactions induced by heavy ions has remained a significant challenge for nuclear physicists for the past five decades. The underlying dynamics governing the fusion of two heavy ions and the subsequent evolution of the fused system manifest themselves through intriguing features over a wide range of beam energy.  Experimentally, it has been well established that sub-barrier fusion cross-sections are much more enhanced than predictions of one-dimensional barrier penetration calculations. This is now well understood as due to the coupling between different channels \cite{dasgupta_1998}. Interestingly, as one goes down much below the barrier, the couplings seem to vanish leading to lower-than-expected fusion cross-sections, a feature, now known as deep sub-barrier fusion hindrance \cite{dasgupta_2007, jiang_2002}. The studies in the fusion of heavy ions continue well above the barrier to understand the response of the hot and rotating nucleus with angular momentum and temperature. The fused nucleus may evaporate light particles and stabilize to form an evaporation residue (ER) or may undergo fission into two massive fragments of two equal or unequal masses. The measurement of the survival probability of the compound system against fission is again a topic of great interest. Experimentally, one measures either the evaporation residues or the fission fragments and compares them with statistical or dynamical model calculations. The sum of the fission and ER cross-sections tells us about the total fusion cross-section. The  substantial amount of experimental data and phenomenological analyses have established that the cross-section of the formation of the compound system is intricately connected with the beam energy and the target-projectile combination in the entrance channel. The mass asymmetry in the target-projectile system is particularly crucial in determining the fate of the system. The target and projectile may mutually get captured to form a di-nuclear system, to be separated again as target- and projectile-like fragments, a phenomenon known as quasi-fission \cite{back_1983,sagaidak_2003,sagaidak_2022,nature_2001, ANU_2021}. In addition, one may also encounter the process known as fast fission where the mono-nuclear system instantly breaks into two nearly equal masses. This process is likely to happen at a critical angular momentum where the potential pocket vanishes and the mono-nucleus rolls back to the scission point. In the case of complete equilibration in all degrees of freedom, a compound system is formed which may either produce the ER or may undergo fission \cite{Ngo_1983}. In addition to the entrance channel mass asymmetry, the effect of shell structure has also been a topic of much research. ER cross-sections have been measured and analyzed in light of the closed shell structure of the target or the compound system \cite{Satou_2002,vsingh_2014,gayatri_2013}. The effect of target and projectile deformations governing the fusion process has also been a topic of much interest \cite{back_1980}. In order to have a better insight into the rather complex nature of the fusion-fission process, it is necessary to carry out exclusive measurements of the ER at different beam energies and angular momentum windows. In addition, it is also necessary to measure the ER from the same compound nucleus produced by different target-projectile combinations with differing mass-asymmetry parameters in the entrance channel. \\ \hspace{31cm} In the present experiment, we have measured ER cross-sections and ER-gated gamma-ray fold distributions for the $^{186}$Pt compound nucleus populated by bombarding $^{32}$S beam on $^{154}$Sm target. 
The ER cross-sections for this system have previously been measured by Gomes \textit{et al.} \cite{gomes_1994} primarily below and around the barrier. The fission fragments for the same system had earlier been measured by Back \textit{et al.} \cite{back_1980, glagola_1984}.  
  The present measurements have added five new data points on the ER cross-sections above the barrier and have also generated the spin distributions for the first time. The ER cross-sections for the same $^{186}$Pt CN have previously been measured by Rajesh \textit{et al.} \cite{rajesh_2019} using a significantly different target projectile combination. An important outcome of the present measurement is the very different ER cross-section from that of \cite{rajesh_2019} over the same region of energies in the center of mass frame. The difference is a clear manifestation of the entrance channel effect on the fusion and subsequently ER production cross-section. The measured ER cross-sections and the spin distributions have been analyzed within the framework of both statistical model and the dynamical DNS model.
The present analysis of the ER cross-section using the DNS model also hints toward the role of incomplete fusion (ICF). In addition to the DNS calculations, we have tried to analyze the ER data using a statistical model and have examined the roles of variable fission barriers and nuclear viscosity. We have generated the range of the parameter spaces for both the barrier and the Kramers' viscosity parameter that reproduces the data.  
\\ The paper is organized into five sections. Section \ref{sec:level2} provides the experimental details followed by a detailed description of the data analysis in Section \ref{sec:level3}. The sub-sections \ref{HYRA_eff1}, \ref{ER-cross-section}, and \ref{ER gated spin distribution} in Section \ref{sec:level3} discuss the determination of HYRA efficiency, ER cross-sections, and ER-gated spin distributions, respectively. Theoretical calculations are presented in Sections \ref{sec:level4} with the statistical model and dynamical model calculations given in sub-sections \ref{statistical_model} and \ref{DNS_model}, respectively. The results obtained from the analysis are discussed in section \ref{sec:level5} followed by a summary in Section \ref{sec:level6}.

\section{\label{sec:level2}Experimental Details}
\begin{figure}[h!]
    \centering
   \includegraphics[width=8.5cm]{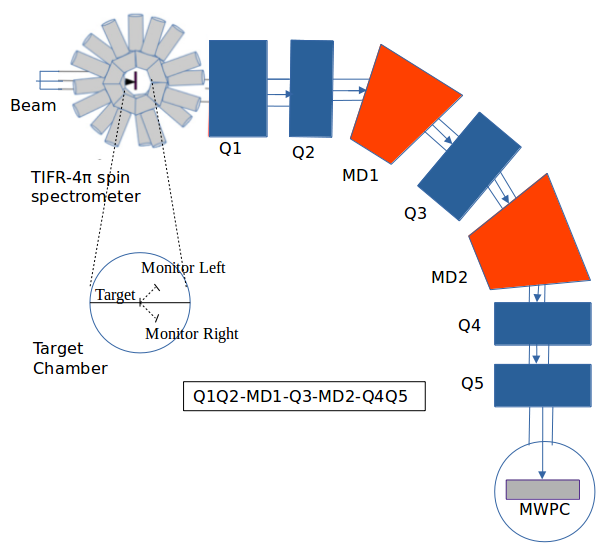}   
\caption{(Color online) A schematic diagram (not to the scale) of TIFR 4$\pi$ Sum-Spin spectrometer and gas-filled HYbrid Recoil mass Analyzer (HYRA). Q1-Q5 are the magnetic quadrupoles;
MD1 and MD2 are magnetic dipoles. MWPC is the Multi-Wire Proportional Counter at the focal plane of HYRA.}
    \label{fig_1}
    \end{figure}

The experiment was performed at the Inter-University Accelerator Centre (IUAC), New Delhi \cite{iuac}. The $^{186}$Pt compound nucleus was populated by bombarding the $^{154}$Sm target with a pulsed beam of $^{32}$S from the Pelletron - LINAC accelerator facility at IUAC. An average beam current of $\sim$ 0.5 pnA was maintained throughout the experiment. The 98.89\% isotopically enriched $^{154}$Sm target of thickness 118 $\mu$g/cm$^2$ had carbon capping and backing of 10 $\mu$g/cm$^{2}$ and 25 $\mu$g/cm$^{2}$, respectively. The measurements were performed at lab energies (E$_{\rm lab}$) of 148.4, 154.8, 176.4, 181.3, 186.4 and 191.5 MeV with pulse separation of 2 $\mu$s. The detection facility involved the HYbrid Recoil mass Analyzer (HYRA) in gas-filled mode, coupled with the TIFR 4$\pi$ sum-spin spectrometer. The configuration of the electromagnetic elements of gas-filled HYRA is Q1Q2-MD1-Q3-MD2-Q4Q5 where Q and MD stand for quadrupole and magnetic dipole, respectively. The schematic diagram of the setup is shown in Fig. \ref{fig_1}. For operating HYRA in gas-filled mode, Helium gas was used at a pressure of 0.21 Torr. A carbon foil with a thickness of 650 $\mu$g/cm$^2$ was used to segregate the gas-filled zone from the beam line maintained in vacuum. For a fuller description and technical details of HYRA, we refer to \cite{madh_2010}. The optimization of the magnetic field values for different beam energies was done using a simulation program \cite{nath_mag}. The optimum field values were chosen by scanning within $\pm$ 10\% of the calculated values in steps of 2\% for each lab energy.  The ERs, filtered through the mass analyzer, were detected in a multi-wire proportional counter (MWPC) at the focal plane of HYRA. The MWPC had an active area of 150 mm × 50 mm and was operated using isobutane gas at a pressure of $\sim$ 2.0 Torr. A 0.5 $\mu$m thick mylar foil was used to separate the MWPC from the gas-filled region of HYRA. Two silicon surface barrier detectors (referred to as monitor detectors in the text) were used to measure the Rutherford (elastically) scattered beam for normalizing the reaction cross-sections. These detectors were placed inside the target chamber at a distance of 47 mm from the center of the target and at an angle of 23.4$^{\circ}$ polar angle in the horizontal plane. The spin distribution was measured by detecting the low energy, discrete $\gamma$-rays using the TIFR-4$\pi$ spin-spectrometer around the scattering chamber. The spin spectrometer is an array of 32 NaI(Tl) detectors in a soccer ball geometry covering 4$\pi$ solid angle. The NaI(Tl) detectors are conical in shape with pentagonal and hexagonal cross-sections. The array is comprised of 12 hexagonal and 20 pentagonal detectors. Together they form a spherical shell around the target. In the present measurements, 29 out of 32 detectors were used after accommodating the inlet and outlet sections of the beam pipe and the target ladder. The compact array of the 29 detectors covered a total solid angle of around 86\% of 4$\pi$. Each of the detectors was energy calibrated with  $\gamma$-ray sources of $^{137}$Cs (662 keV) and $^{60}$Co (1173 and 1332 keV). The response of the array to single and multiple $\gamma$-rays was simulated using Geant4 package \cite{anil_2008, anil_2009, asit_2023}. The determination of the fold-to-spin distribution using the spectrometer will be discussed in detail in section V.  The cleanly separated ER events were obtained using the time-of-flight (TOF) approach. We used two time-to-amplitude converters (TACs) for this purpose. The first TAC measured the time difference between the start signal from the MWPC anode and the stop signal from the rf signal of the beam. The second TAC measured the time difference between the start signal from the MWPC-anode and the stop signal from suitably delayed logic OR of all the NaI(Tl) detectors in the 4$\pi$ spin spectrometer. The master strobe for data acquisition was the logic OR signal of the two monitor detectors and the anode signal of the MWPC. The CANDLE software was used for data collection and reduction \cite{et_candle}.
\section{\label{sec:level3}Data Analysis}This section presents a detailed account of the data analysis conducted for this experimental work. In particular, Subsection \ref{HYRA_eff1} describes the procedure for calculating HYRA efficiency, which is critical for estimating ER cross-sections. Furthermore, Subsection \ref{ER-cross-section} explains the methodology used for determining experimental ER cross-sections. The analysis of the ER-gated spin distribution is provided in Subsection \ref{ER gated spin distribution}.
\subsection{\label{HYRA_eff1}Determination of HYRA efficiency}
The transmission efficiency of HYRA is one of the important factors for the determination of the ER cross-sections. It can be defined as the ratio of the total ERs detected at the focal plane  to the total ERs produced in the target and depends on the beam energy, entrance channel, target thickness, exit  channel, magnetic field values, the angular acceptance of HYRA, gas pressure settings of the HYRA, and the size of the focal plane detector (MWPC) \cite{snath_2010}. To find the transmission efficiency (${\epsilon_{H}}$) for the present measurement, we have used the formula:
\begin{equation}
     \epsilon_{H} = \frac{Y_{ER}}{Y_{Mon}} 
    \left( \frac{d\sigma}{d\Omega}\right)_{Ruth}\Omega_{Mon}\frac{1}{\sigma_{ER}}
    \label{equation_1}
    \end{equation}
    where $Y_{ER}$ is evaporation residue yield at the focal plane, $Y_{Mon}$ is the yield of elastically scattered projectiles detected by silicon surface barrier detectors and given by the geometric mean ($\sqrt{(Y_L Y_R)}$) of counts in thetwo detectors placed at 23.4$^{\circ}$, $\left( \frac{d\sigma}{d\Omega}\right)_{Ruth}$ is the differential Rutherford scattering cross-section, ${\rm {\Omega_{Mon}}}$ is the solid angle subtended by monitor detector at the center of the target, ${\sigma_{ER}}$ is ER cross-section.
  The differential Rutherford scattering is given by, 
\begin{equation}
\left( \frac{d\sigma}{d\Omega}\right)_{Ruth} \approx 1.296 \left(\frac{Z_p Z_t}{E_{lab}}      \right)^{2}\left[\frac{1}{{\rm sin}^4{(\frac{\theta}{2})}}- 2\left(\frac{A_p}{A_t}\right)^2\right]
\label{equation3}
        \end{equation}
 where $Z_p$ and $A_p$, $Z_t$ and $A_t$ are the atomic and mass numbers of the projectile and target, respectively. $E_{\rm lab}$ and $\theta$ are the energy of the incident projectile in the lab frame and the scattering angle of the projectile-like particles in the laboratory frame of reference, respectively.
    Experimental ER cross-sections for the $^{32}$S +$^{154}$Sm system at lower energies than the present measurements are already available in literature \cite{gomes_1994}. We have matched one energy point $E_{\rm c.m.}$=127.9 MeV at the center of the target ($E_{\rm c.m.}$=128.4 MeV just before the target) with the present measurements to find the efficiency of HYRA.   
      \begin{figure}[h!]
    \centering
   \includegraphics[width=8.5cm]{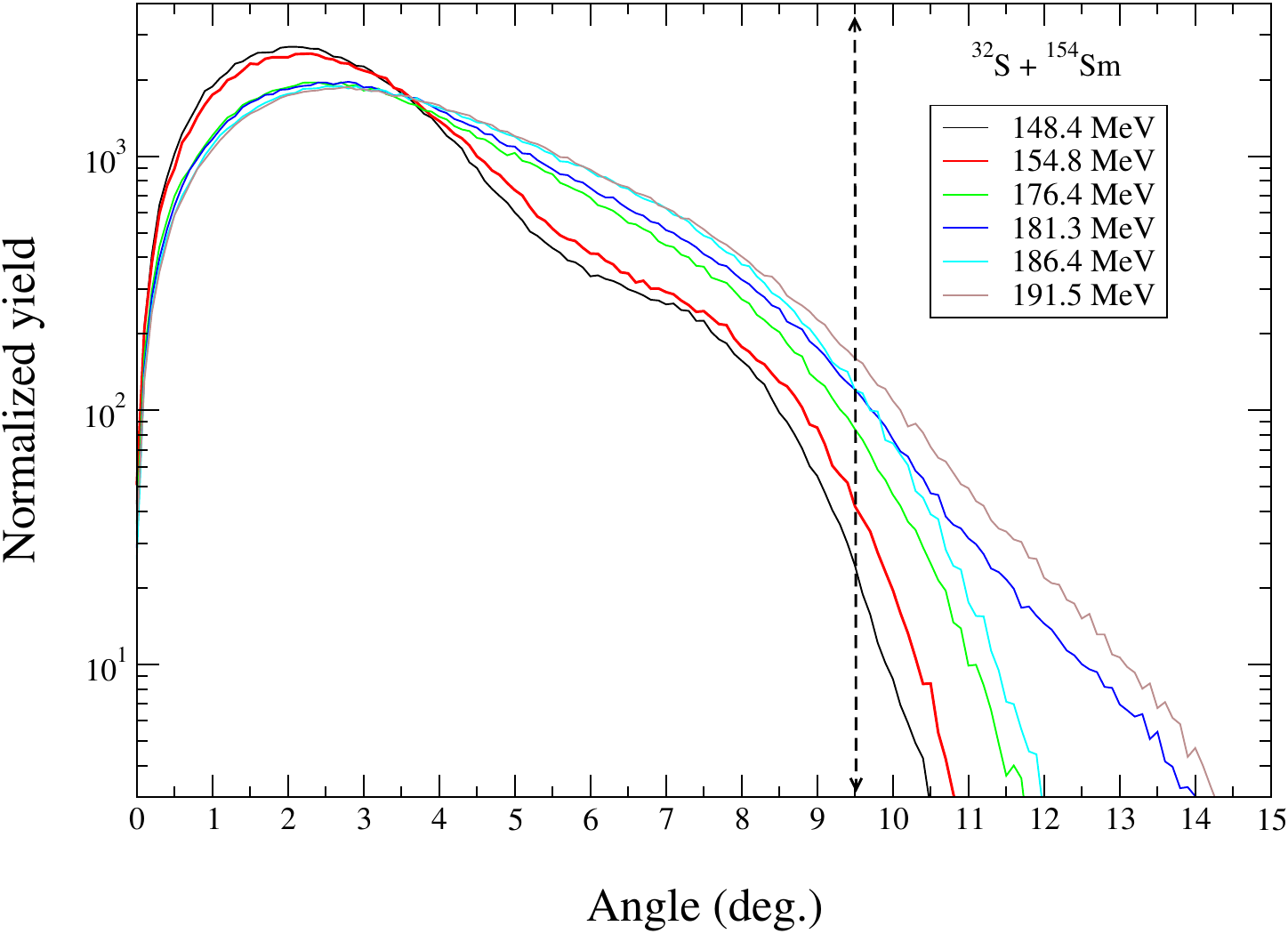}   
    \caption{(Color online) Normalized angular distributions of ERs for the $^{32}$S +$^{154}$Sm at different lab energies calculated using TERS code \cite{nath_2009}. Angular acceptance of HYRA is 9.5$^\circ$ (marked as a vertical dashed line).}
    \label{fig_2}
\end{figure}The experimentally extracted efficiency was found to be 10.7\%. The transmission efficiencies of HYRA at other energies were obtained by multiplying $\epsilon_{H}$ at energy 127.9 MeV with the ratio of the  area  under the curve of the normalized ER angular distribution yield at 127.9 MeV to the energy at which efficiency to be evaluated. Angular distribution of ERs was calculated using the semi-microscopic Monte Carlo simulation code TERS \cite{nath_2009} and the relative yield of each ER channel for different energies was calculated using the statistical model code PACE4 \cite{pace4}. TERS calculates the event-by-event interaction of the beam with the target and generates the ER angular distribution spectrum by taking realistic input like neutron, proton and $\alpha$ separation energies. The charge-state acceptance of HYRA was assumed to be 100\%, but due to the restricted aperture of the target chamber used, the acceptance angle was limited to 9.5$^\circ$. Thus, only the area under the curve up to 9.5$^\circ$ was considered. The suitable weight factor was multiplied with individual ER channels, then adding all ER channels resulted in normalized yield at each energy. The angular distributions of the ER yield for different lab energies are shown in Fig. \ref{fig_2}. The extracted values of transmission efficiencies are shown in Table \ref{table1}.  
 \begin{table}
\centering
\caption{\label{table1}Extracted transmission efficiencies for ERs at different beam energies for the $^{32}$S +$^{154}$Sm reaction.\\}
\begin{tabular}{l l c  }
 \hline
\hline
$E_{\rm lab}$ (MeV) & $E_{\rm c.m.}$ (MeV) & Transmission efficiency ($\epsilon_H$) ($\%$) \\[1ex] 
\hline \\
148.4  & 122.9  & 10.7 $\pm$  2.1         \\ [1ex] 
154.8  & 128.2  & 10.7 $\pm$  2.0         \\ [1ex]
176.4  & 146.0  & 11.4 $\pm$  2.1         \\ [1ex]
181.3  & 150.1  & 11.0 $\pm$  2.1         \\ [1ex]
186.4  & 154.4  & 10.8 $\pm$  2.1         \\ [1ex]
191.5  & 158.6  & 10.8 $\pm$  2.1         \\ [1ex]
\hline
\hline
\end{tabular}

\end{table}

\subsection{Determination of ER cross-sections}\label{ER-cross-section}

Evaporation residue cross-sections were calculated using the formula: 
\begin{equation}
    \sigma_{ER} = \frac{Y_{ER}}{Y_{Mon}} 
    \left( \frac{d\sigma}{d\Omega}\right)_{Ruth}\Omega_{Mon}\frac{1}{\epsilon_{H}}
    \label{equation_2}
    \end{equation}

        \begin{figure}[h!]
    \centering
      \includegraphics[width=8.7cm]{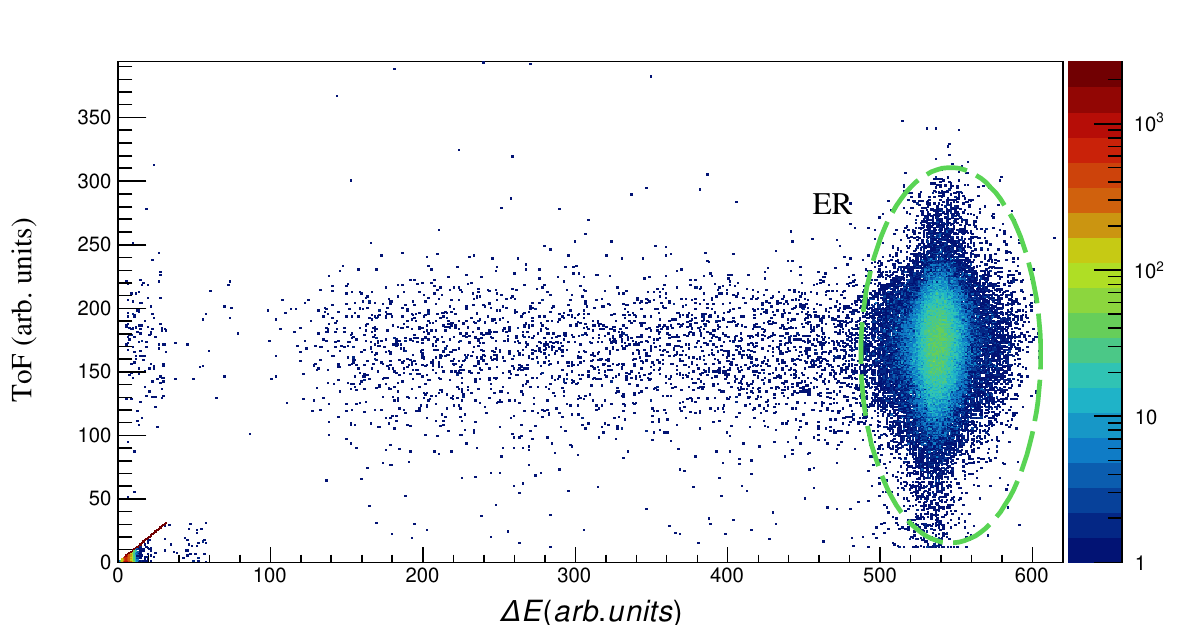}     
            
        \caption{(Color online) Two-dimensional spectrum between energy loss $(\Delta E)$ signal from the  MWPC cathode and TOF for ERs of $^{186}$Pt compound nucleus at lab energy 181.3 MeV ($E_{\rm c.m.}$=150.1 MeV).}
    \label{fig_3}
\end{figure}

\begin{figure}[h!]
    \centering
   \includegraphics[width=8.5cm]{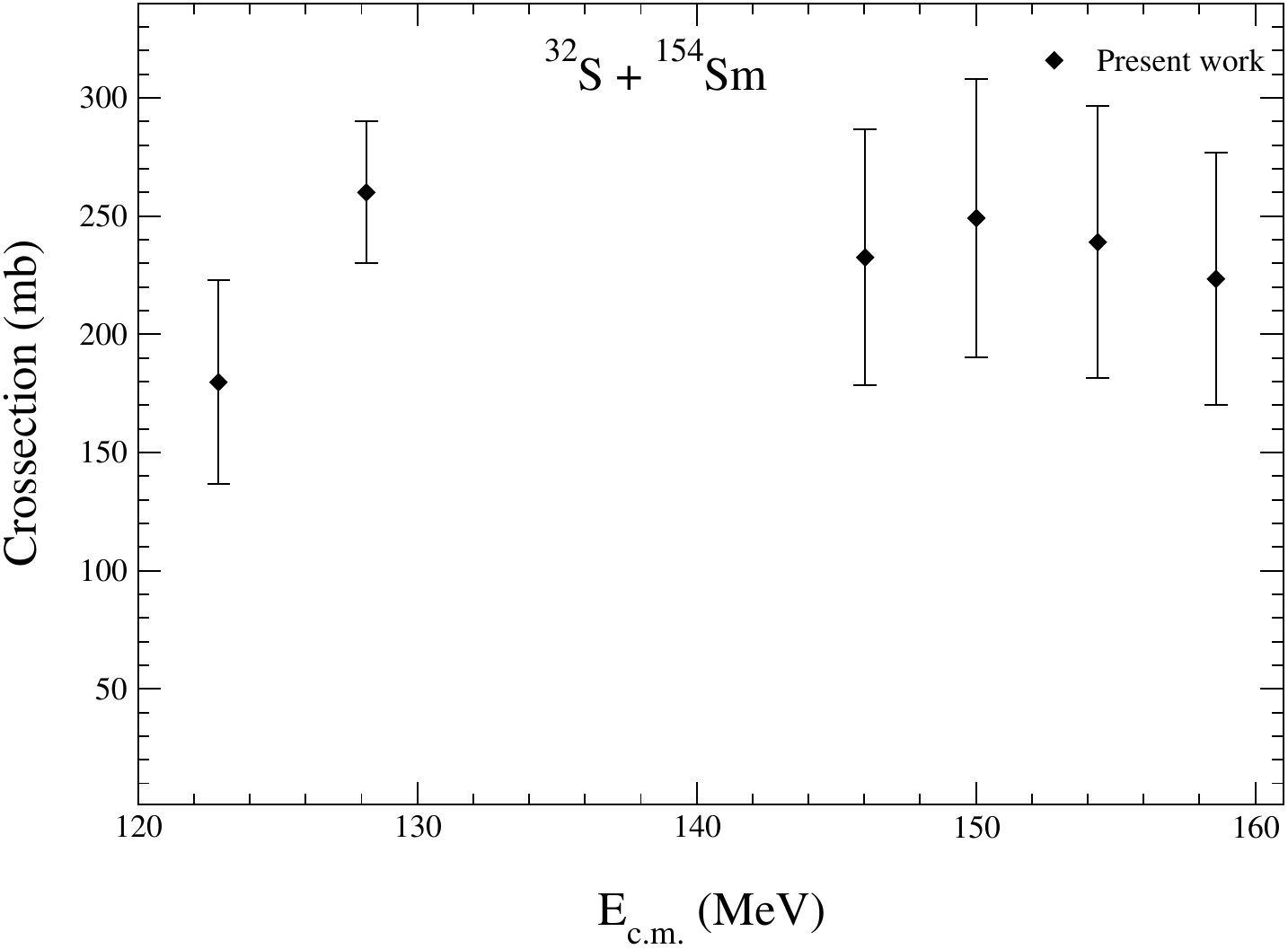}     
             \caption{ The measured ER cross-sections for $^{186}$Pt compound nucleus as a function of the center of mass energy ($E_{\rm c.m.}$). }
    \label{fig_4}
\end{figure}

\begin{figure}[h!]
    \centering
   \includegraphics[width=8.5cm]{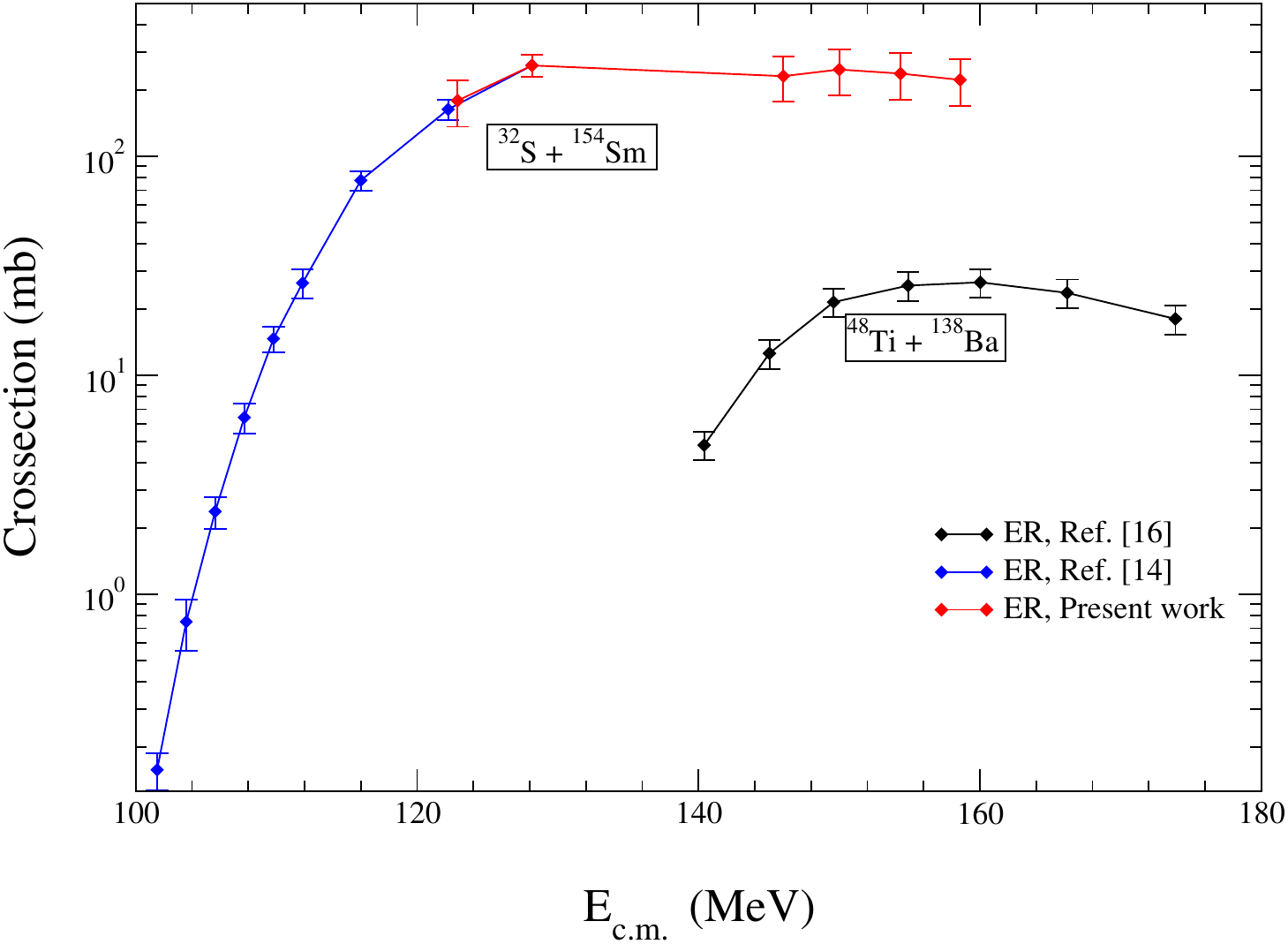}     
        \caption{ (Color online) Comparison of ER cross-sections of $^{186}$Pt compound nucleus as a function of the center of mass energy ($E_{\rm c.m.}$) for two different entrance channels (target-projectile combinations) as discussed in the text. (Lines are to guide the eyes only)}
    \label{fig_17}
\end{figure}

\begin{table}
\centering
\caption{\label{table2}Total ER cross-section of $^{32}$S +$^{154}$Sm at different energies.\\}
\begin{tabular}{ l c c c   }
\hline
\hline
\vspace{0.1cm}
$E_{\rm lab}$ &  $E_{\rm c.m.}$& E$^{*}_{CN}$ & $\sigma_{ER}$ $\pm$ error\\[1ex] 
(MeV) & (MeV) & (MeV) & (mb) \\[1ex]
\hline \\
148.4 & 122.9 & 62.3 &  180   $\pm$ 43
         \\ [1ex] 
154.8 & 128.2 & 67.6 &  260      $\pm$    30
      \\ [1ex]
176.4 & 146.0 & 85.4 &  232   $\pm$      54
    \\ [1ex]
181.3 & 150.1 & 89.5 &  249    $\pm$  59
        \\ [1ex]
186.4 & 154.4 & 93.8 &  239       $\pm$   58
        \\ [1ex]
191.5 & 158.6 & 98.0 &  223   $\pm$          53      \\ [1ex]
\hline
\hline
\end{tabular}

\end{table}
     Fig. \ref{fig_3} shows a typical 2D-spectrum showing the ERs from the compound nucleus $^{186}$Pt at lab energy 181.3 MeV. Total ER cross-section values are given in Table \ref{table2}. ER cross-sections measured in the present experiment are shown in Fig. \ref{fig_4}. The errors in the experimental cross-section (as shown in Table \ref{table2}) were calculated using error in the HYRA efficiency ($\rm{\epsilon_{H}}$), statistical error obtained from the measured yields $Y_{ER}$ and $Y_{Mon}$ and other systematic errors. The main contribution to the error comes from the HYRA efficiency parameter. 
     Before proceeding further with the next part of the data analysis, namely, the determination of the spin distribution and subsequent theoretical calculations, we would like to point out an  important observation that emerges while comparing our ER cross-sections with a previous set of data. The ER cross-sections from the $^{186}$Pt compound nucleus have been measured in recent times by Rajesh \textit{et al.} \cite{rajesh_2019} over the same energy range covered in the present measurements. However, they have used $^{48}$Ti + $^{138}$Ba reaction. The difference in mass asymmetry is considerable between the two reactions and is reflected in the rather low ER cross-sections reported by Rajesh \textit{et al.} compared to the present measurements. Fig. \ref{fig_17} shows the ER cross-sections from $^{32}$S + $^{154}$Sm and $^{48}$Ti + $^{138}$Ba reactions. The very large difference between the cross-sections from the two reactions is a clear demonstration of the effect of mass asymmetry in the entrance channel on the formation of the compound nucleus and eventual ER cross-sections.  Theoretical models have suggested fusion inhibition for projectile and target combinations with $Z_p$$Z_t$ $\geq$ 1600 \cite{blocki_1986, swiatecki_1981}. However, in the compound nucleus $^{216}$Ra populated through different entrance channels, the fusion has been found to be suppressed down to approximately half of the predicted value of $Z_p$$Z_t$ \cite{nature_2001}. This reduction in fusion is ascribed to the quasi-fission resulting from mass asymmetry in the entrance channels, which causes a decrease in the cross-section of evaporation residues. In light of this, the lower ER cross-sections reported by Rajesh \textit{et al.} \cite{rajesh_2019} compared to the present work can be attributed to the difference in mass asymmetries in the two entrance channels.  

\subsection{Determination of ER gated spin distribution}\label{ER gated spin distribution}
\begin{figure}[h!]
    \centering
\includegraphics[width=8.5cm]{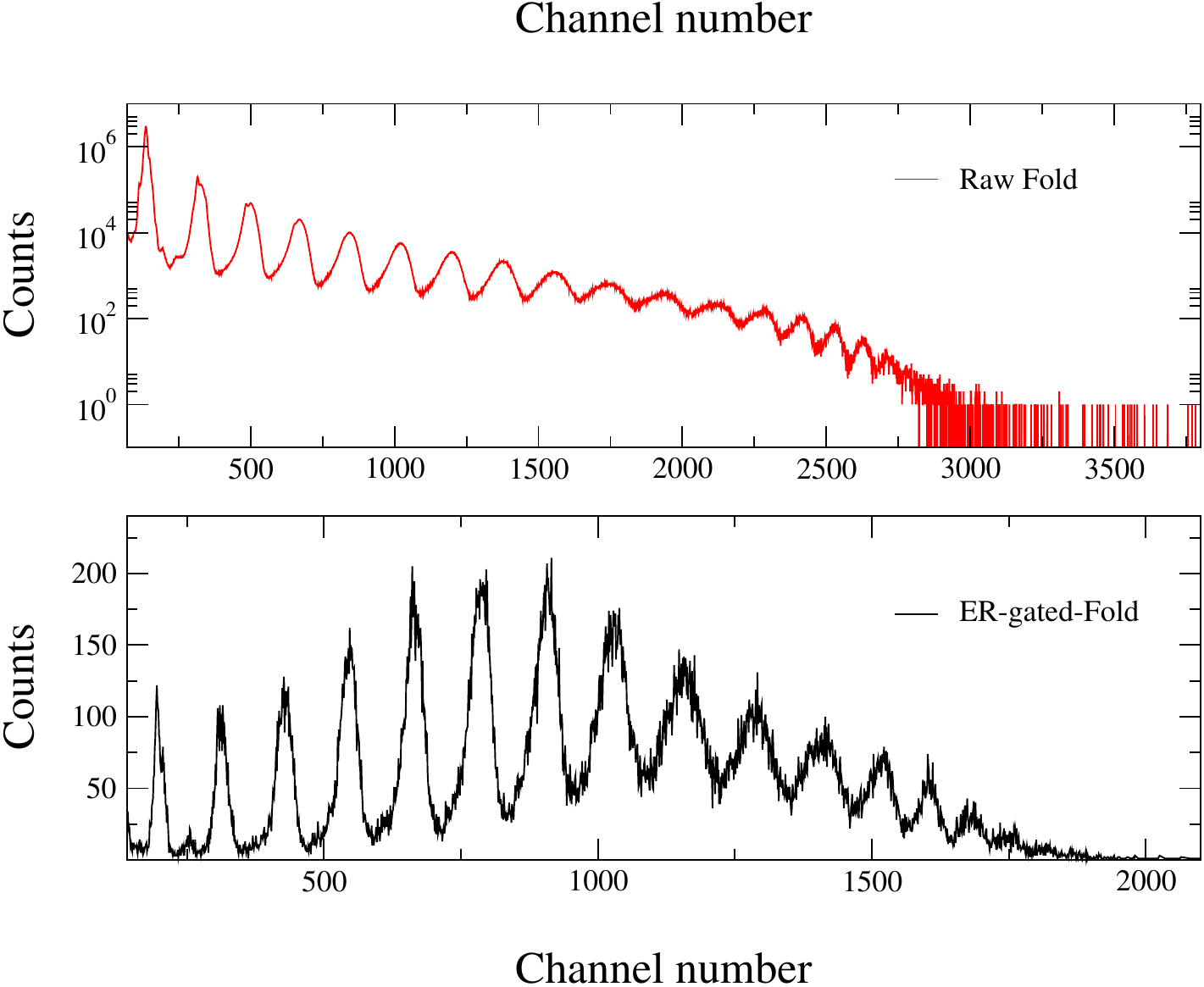}
\caption{(Color online) Raw and ER-gated $\gamma$-ray fold distribution at $E_{\rm lab}$ = 191.4 MeV.}
    \label{fig_5}
\end{figure}

The ER spin distribution was measured using the TIFR 4$\pi$ spin spectrometer. The spin spectrometer has been described in some detail in section \ref{sec:level2}. We also refer to previous measurements with the HYRA + 4$\pi$ spin spectrometer combination \cite{gayatri_2012, gayatri_2013, priya_2017,sudarshan_2017,muthu_2020}.
All 29 detectors used in the present measurements were time-aligned using homemade delay units. The energy threshold of each detector was set at about 120 keV. Timing OR and multiplicity signals from all detectors were generated using multi-channel CFD (Constant Fraction Discrimination) units. Fig. \ref{fig_5} presents typical raw and ER-gated fold distributions for the reaction at 191.4 MeV beam energy.  The difference in the two spectra clearly demonstrates the power of the ER-tagging using HYRA. 
The next steps in the analysis involved extraction of the spin distribution from the experimentally measured fold distributions for the different beam energies. Fold distribution  probability $P(k)$ can be given by:\begin{equation}
P(k) = \sum_{M_{\gamma=0}}^{\infty}R(k,M_{\gamma}) P(M_{\gamma})
\label{equation_4}
        \end{equation}
where $R(k, M_{\gamma})$ is the response function, in other words, it is the probability of firing $k$ detectors out of $N$ detectors for $M$ uncorrelated $\gamma$ rays and $P(M_{\gamma})$ is the probability of multiplicity distribution. 
\begin{figure}[h!]
    \centering
\includegraphics[width=8.5cm]{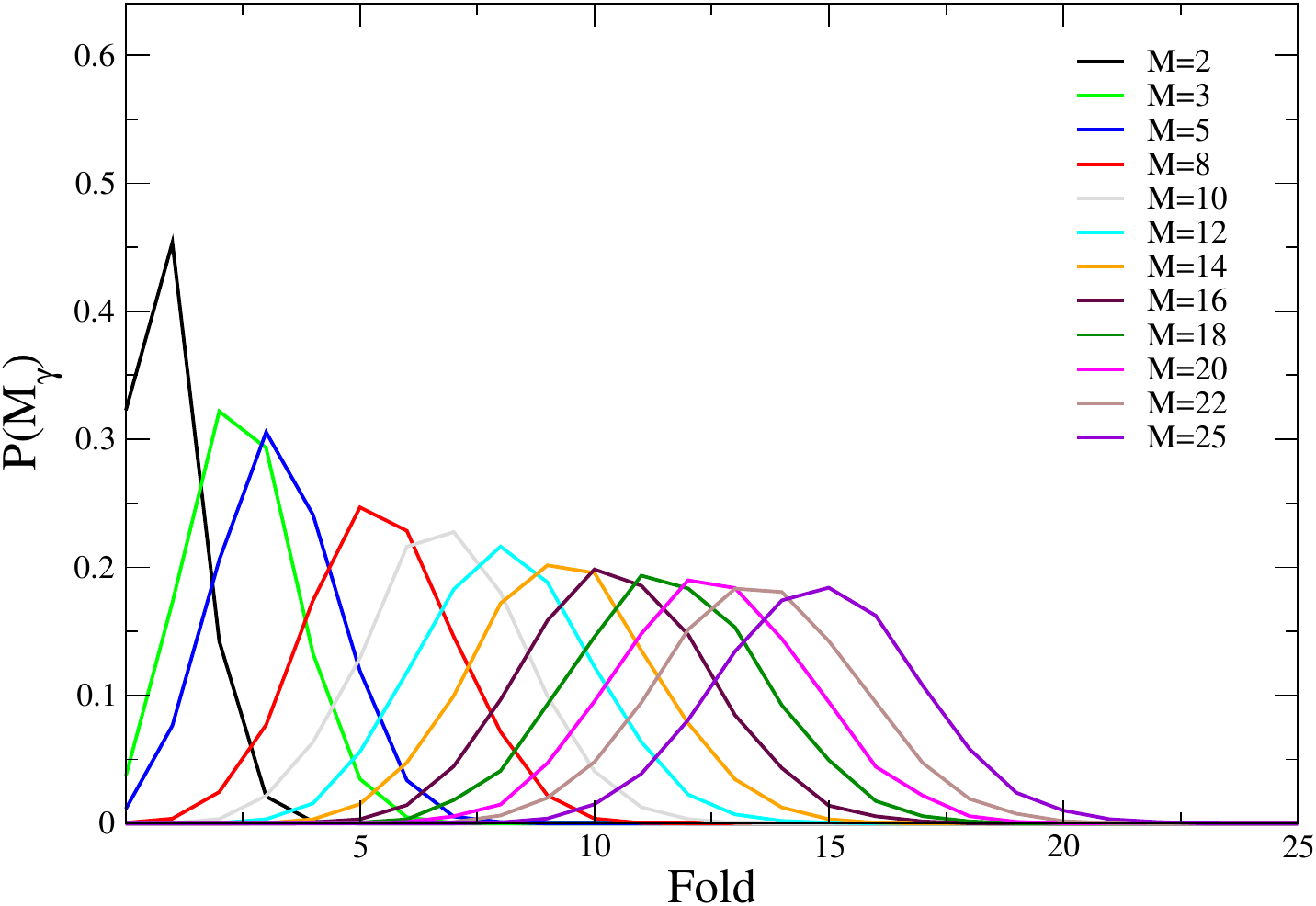}
\caption{(Color online) Geant4 \cite{geant_4,geant4} simulated response of TIFR 4$\pi$ spin spectrometer for different $\gamma$-ray multiplicities (M$_\gamma$) events. }
    \label{fig_6}
\end{figure}
We have carried out realistic simulations of the response of the spectrometer to multiple gamma-rays using the Geant4 package \cite{anil_2008, anil_2009, asit_2023,geant_4,geant4}. Fold distributions for different multiplicities, i.e. for a given gamma multiplicity $M$, distribution in fold $k$ for the spin-spectrometer were generated.
The Geant4 simulated response of TIFR 4$\pi$ spin spectrometer for different gamma multiplicities $M_{\gamma}$ is shown in Fig. \ref{fig_6}. Traditionally, one convolutes the response function with the multiplicity distribution of some functional form and fits the experimentally measured fold data with free parameters \cite{gayatri_2013,muthu_2020,Werf_1978,shidling_2006,nath_2011}. The assumed multiplicity distribution with the Fermi function form is given by:   
\begin{equation}
P(M_\gamma) = \frac{2M_\gamma+1}{\rm{exp}(\frac{M_\gamma-M_o}{\Delta M})+1}
\label{equation_fermi}
        \end{equation}
where $M_\gamma$ is gamma-multiplicity, $M_o$ and $\Delta M$ are free parameters. The fold distribution is then simulated by convoluting the response function $R(k,M_{\gamma})$ with the Fermi function as given in Eq. (\ref{equation_fermi}) and by varying free parameters $M_o$ and $\Delta M$,
to fit the experimental fold data. The best fit to the experimental fold distribution with free parameters gives the most probable spin distribution at a particular energy. Fig. \ref{fig_7} shows the simulated fold fitted to the experimental fold for $E_{\rm lab}$=154.8 MeV with free parameters $M_0$ =10 and $\Delta M$=2.
\begin{figure}[h!]
    \centering
   \includegraphics[width=8.5cm]{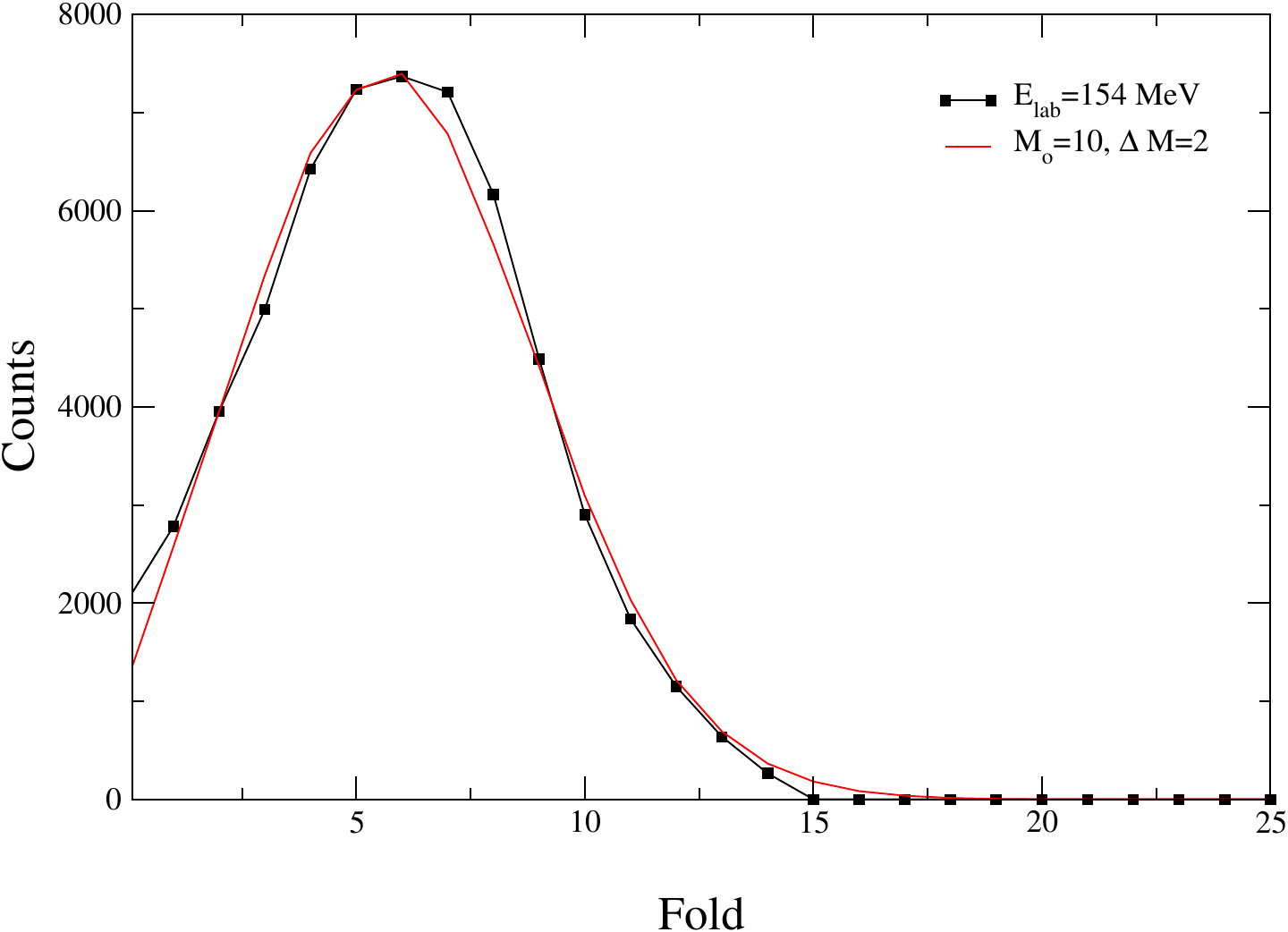}     
    
        \caption{(Color online) The experimental and simulated fold distributions for $^{32}$S +$^{154}$Sm at $E_{\rm lab}$=154.8 MeV.}
    \label{fig_7}
\end{figure}
\begin{figure}[h!]
    \centering
   \includegraphics[width=8.5cm]{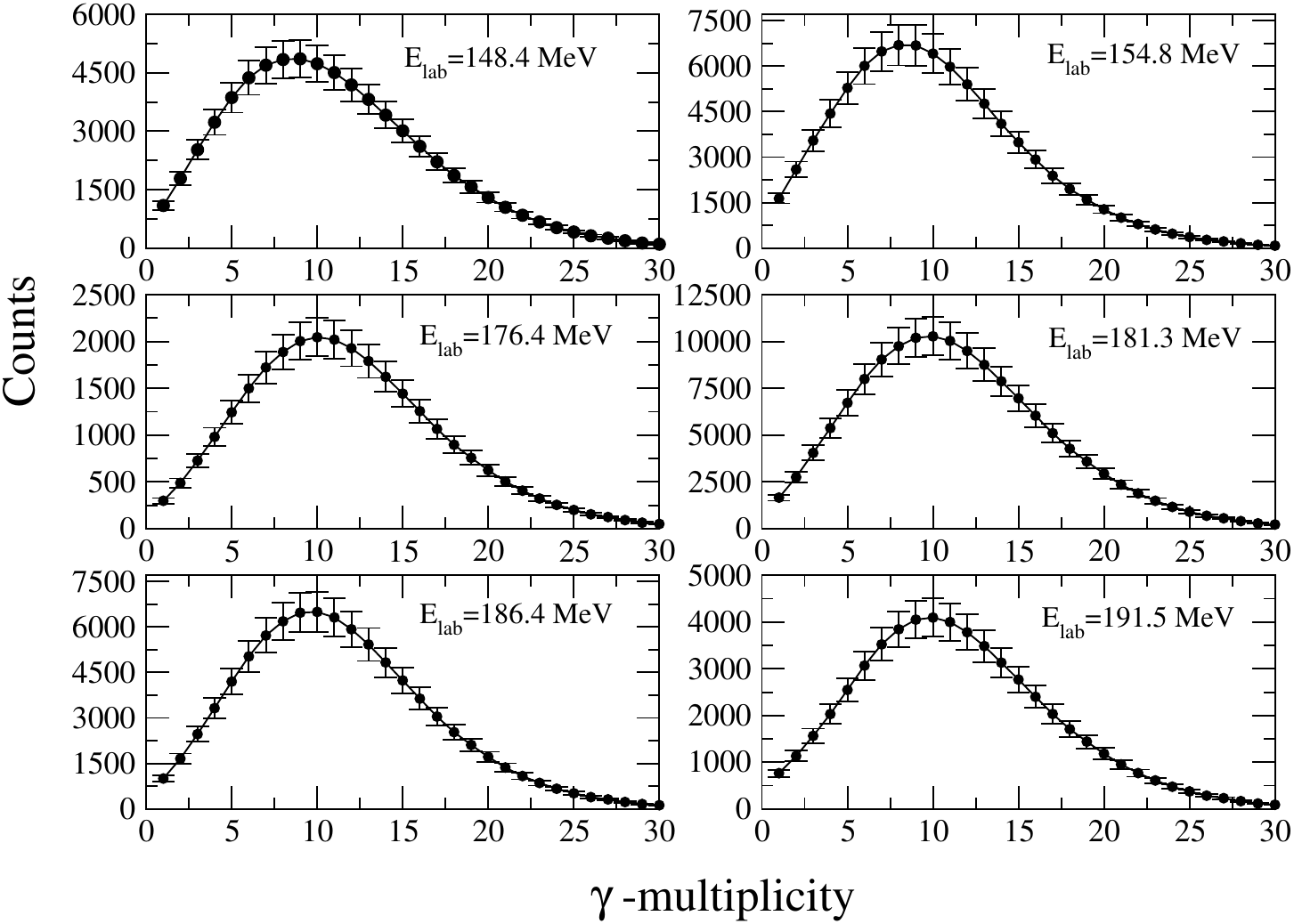}     
    \caption{Extracted $\gamma$-multiplicity distributions associated with ER formation for the $^{32}$S +$^{154}$Sm system at different $E_{\rm lab}$.}
    \label{fig_8}
\end{figure}
\par We have also tried to extract the multiplicity and spin distributions directly from the experimental fold distribution without assuming any specific functional form for the spin distribution. Here again, we make use of the response function $R(k, M_{\gamma})$ generated by the Geant4 simulation. The response function is basically a matrix of given dimensions with each column corresponding to the probability of fold distribution for a particular multiplicity. 
\begin{figure}[h!]
    \centering
  \includegraphics[width=8.5cm]{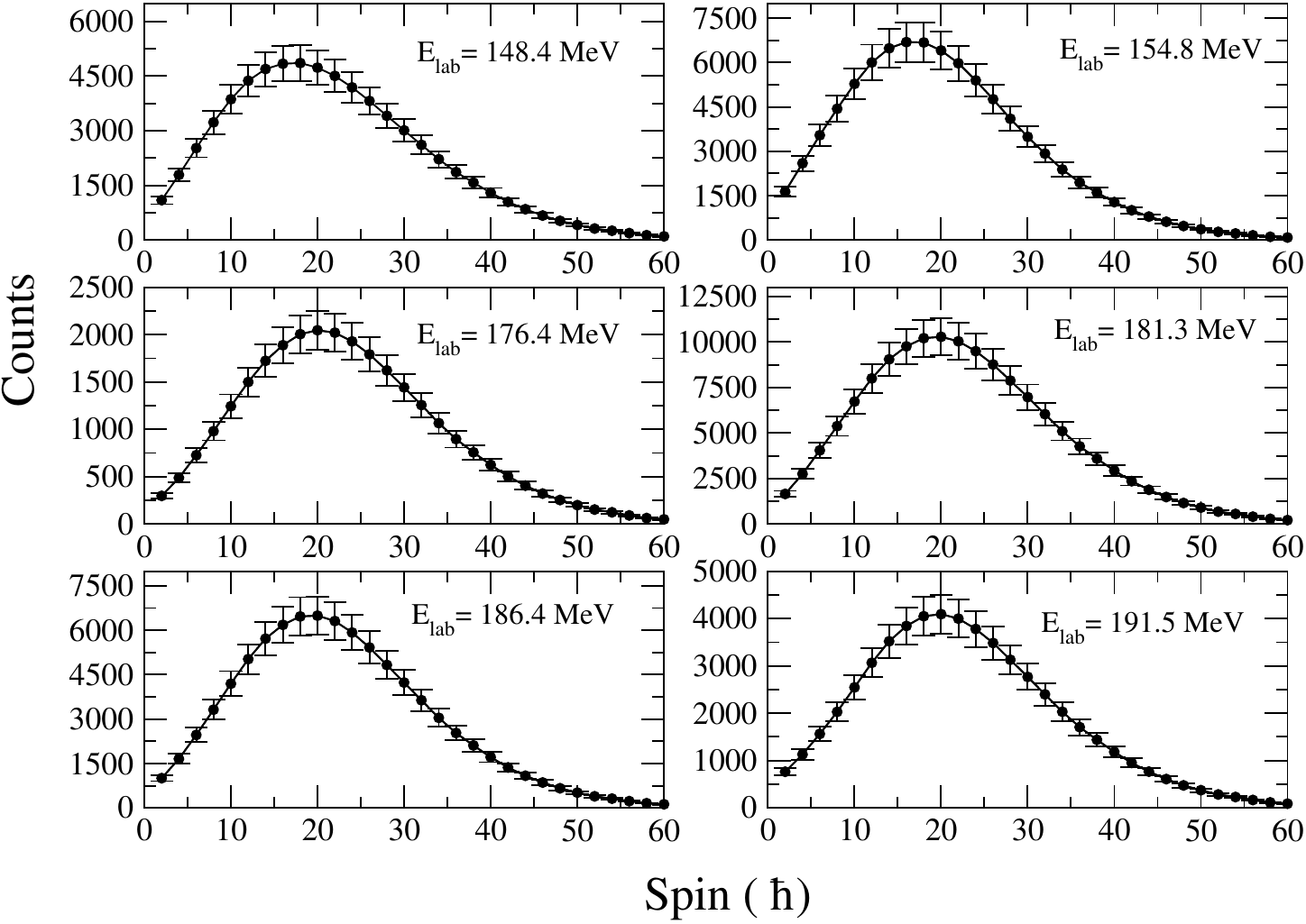}
    \caption{Extracted spin distributions associated with ER formation for the $^{32}$S +$^{154}$Sm system at different $E_{\rm lab}$.}
    \label{fig_9}
\end{figure}
Having carried out the necessary de-convolution process we generate the multiplicity distribution as shown in Fig. \ref{fig_8}. Finally, the spin distribution is generated from the extracted multiplicity distributions for all the beam energies simply assuming each gamma ray carries 2$\hbar$ angular momentum as presented in Fig. \ref{fig_9}. Errors in the multiplicity and spin distribution include $\pm$ 10$\%$ error in the efficiency of NaI(Tl) detector \cite{gayatri_2012}. 
The experimentally measured spin distributions, so extracted from the analysis, have been used as input in the present statistical model calculations.
\section{\label{sec:level4}Theoretical Analysis}
This section is devoted to the theoretical analysis of the experimental results obtained in the present work. As mentioned in the introductory section, we have analyzed the decay of the $^{186}$Pt nucleus to form ERs or fission fragments using phenomenological statistical model calculations. In addition, we have also carried out dynamical calculations within the framework of the DNS model to reproduce the data. 
\subsection{\label{statistical_model}Statistical model calculations}
We have carried out detailed statistical model calculations using a modified version of the CASCADE code \cite{mod_cascade,pramana_2015,cascade_2023,2000_dioszegi,2000_shaw}. At the starting point, the calculations require information about the spin distribution of the compound nucleus formed by a given target-projectile combination. In the analysis, the experimentally measured spin distributions have been taken as input in the CASCADE calculations. The total fusion cross-sections, obtained by summing the experimental ER and ﬁssion cross-sections, are fed into the calculations as input and are distributed by the measured spin distributions. The next very important ingredient to be tuned for statistical model calculations is the Nuclear Level Density (NLD), which is one of the most important physical quantities governing the decay of the compound nucleus. In the present calculations, we have used the Ignatyuk-Reisdorf formalism to calculate the NLD \cite{ignatyuk_1975, reisdrof_1981}.
The Ignatyuk ansatz takes care of the shell effects as a function of excitation energy as given in Eq. \ref{nld_3}. 
\begin{equation} \label{nld_3}
  a(E) = \Tilde{a}\left(1 + \frac{f(E)}{E } \delta W\right)
 \end{equation}
 where $$ f(E) = 1-{\rm exp}\left(\frac{-E}{E_d}\right);$$
$\Tilde{a}$ is the asymptotic
or liquid drop level density parameter, $E_d$ is the rate at which effects of the shell disappear, and $\delta W$ is the shell correction taken from the difference between the experimental and liquid drop model masses.\\
We have used the following expansion 
proposed by Reisdorf    \cite{reisdrof_1981} to calculate $\Tilde{a}$ :
\begin{equation}\label{reisdrof}
\Tilde{a} =0.04543r_0^3A + 0.1355r_0^2A^{2/3}B_S + 0.1426r_0A^{1/3}B_K 
\end{equation}
where $A$ is the nuclear mass, $r_0$ is the nuclear radius, $B_s$
and $B_k$ are the surface and curvature terms of the liquid drop
model, respectively \cite{2000_shaw}.
\begin{figure}[h]
 \centering
 \includegraphics[width=8.5cm]{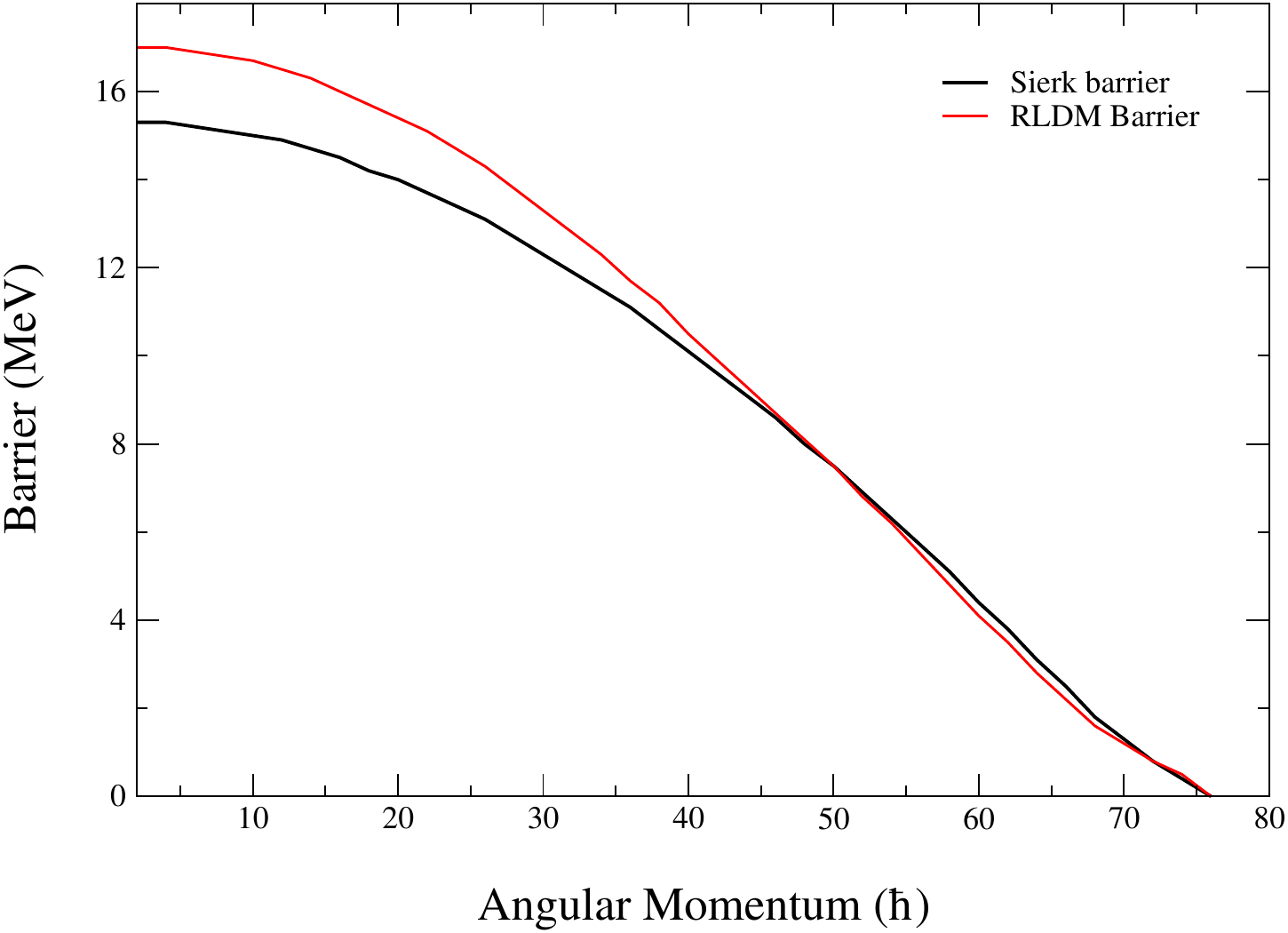}
\caption{(Color online) Sierk and RLDM fission barrier distribution for $^{186}$Pt compound nucleus.}
\label{fig_10}
\end{figure}
 The survival of the compound nucleus against fission and the formation of the ERs are crucially dependent on the fission barrier. The barrier height is dependent on both temperature and angular momentum.
In the present calculations, both the rotating liquid drop model (RLDM) and Sierk (finite-range rotating liquid-drop model (FRLDM)) barriers are used. The Sierk model \cite{sierk_1986} is a more advanced model and differs from the RLDM in many respects. Notable among them are (i) the replacement of the surface energy of the liquid drop model by the Yukawa-plus-exponential nuclear energy and (ii) considering a realistic surface diffusion in the calculation of the Coulomb energy.  The barriers calculated using RLDM and Sierk prescriptions are compared in Fig. \ref{fig_10}. It is seen that the two approaches (RLDM and Sierk) produce significantly different heights at smaller angular momentum. However, as the angular momentum increases the difference narrows down and almost vanishes beyond 40 $\hbar$.\\
In addition to the barrier heights, another physical quantity that controls the flow of the flux from the equilibrium position to the scission point is nuclear viscosity due to dissipative processes. Our modified version of the CASCADE code includes Kramers' prescription to deal with the dissipative mechanism. The entire path from the equilibrium position to the scission point is divided into two parts, namely, equilibrium to the saddle point and then from the saddle to the scission point. The viscosity parameter ($\gamma$) is treated differently for these two regions and is labeled as pre-saddle and post-saddle. For further details, we refer to \cite{pramana_2015}.\begin{figure}[h]
\centering
\includegraphics[width=8.5cm]{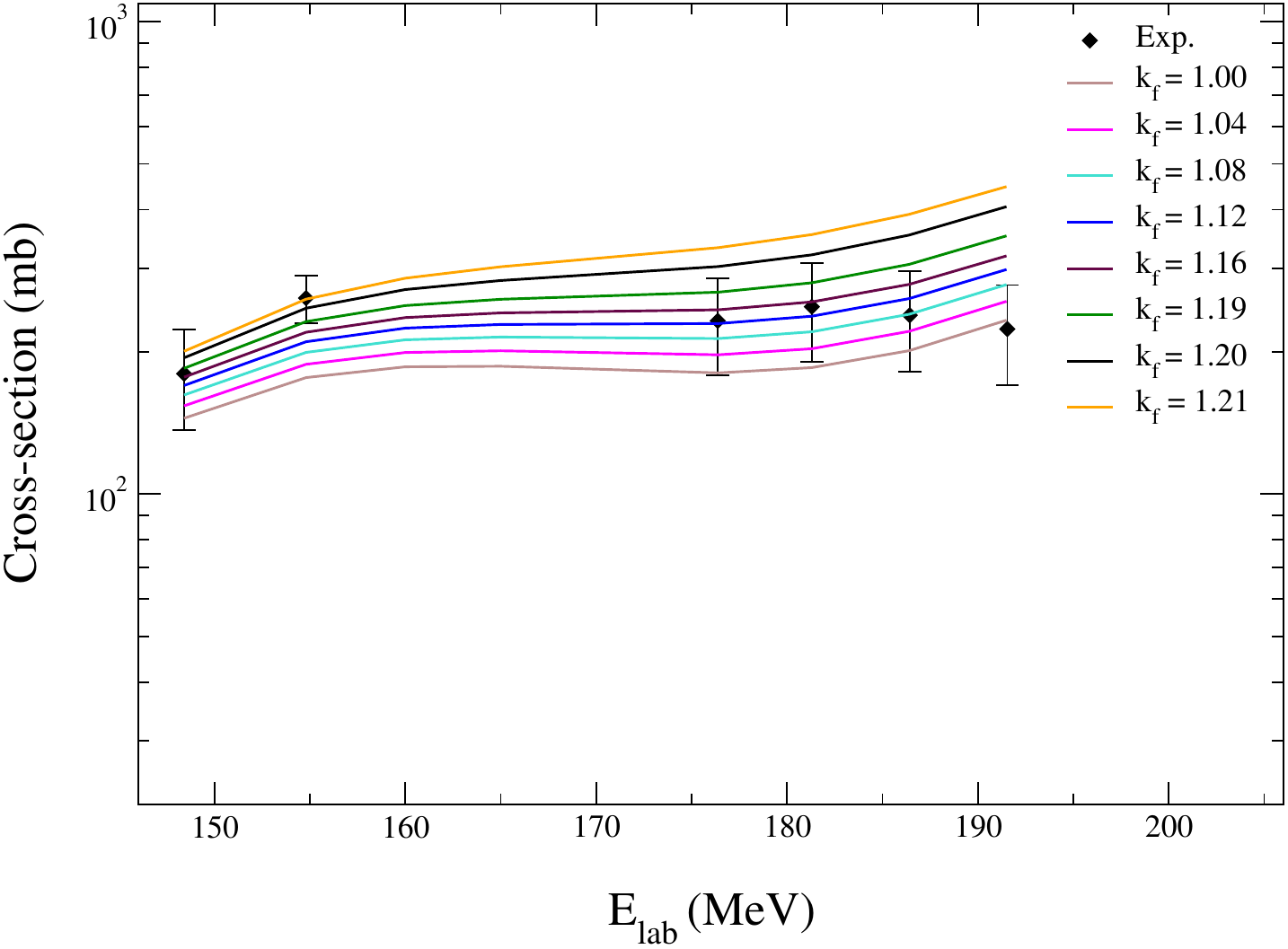}
\caption{(Color online)  Variation of the ER cross-sections for $^{32}$S +$^{154}$Sm with RLDM fission barrier (scaling factor $k_f$) at different $E_{\rm lab}$.}
\label{fig_11}
\end{figure}
The Bohr-Wheeler fission width is reduced by the Kramers' factor as shown below  
  \begin{equation}
\Gamma_f^{Kramers'}=\Gamma_f^{BW}[(1+\gamma^2)^{1/2}-\gamma]
 \label{equation5}
\end{equation}
 where $\Gamma_f^{Kramers}$ is Kramers' fission width and $\Gamma_f^{BW}$ is Bohr-Wheeler fission width. In addition, there is also a temporal variation of the fission width in reaching the Kramers' value as shown below in Eq. (\ref{equation6}).
 \begin{equation}
      \Gamma_f(t)=\Gamma_f^{Kramers}[(1-{\rm exp}(-2.3t/\tau_f)]
         \label{equation6}
 \end{equation} where $\tau_f$ is the delay time for flux to reach quasi-stationary value \cite{1986_grange,1986_bhatt}.
  $ \tau_f=(\beta/2\omega_1^2){\rm {ln}}(10B_f/T)$ where $\beta$=2$\gamma \omega$ is reduced viscosity, $\omega$ is barrier frequency and $B_f$ is fission barrier height. \\
The time taken by the flux to move from the saddle to the scission point is also expected to be prolonged due to the presence of viscosity and is given by \cite{1983_hofmann},
  \begin{equation}
      \tau_{ssc}= \tau^{0}_{ssc}[(1+\gamma^2)^{1/2}+\gamma]
 \label{equation7}
 \end{equation}
 where $\tau^{0}_{ssc}$ is the time without considering any viscosity. 
\begin{figure}[h]
\centering
\includegraphics[width=8.5cm]{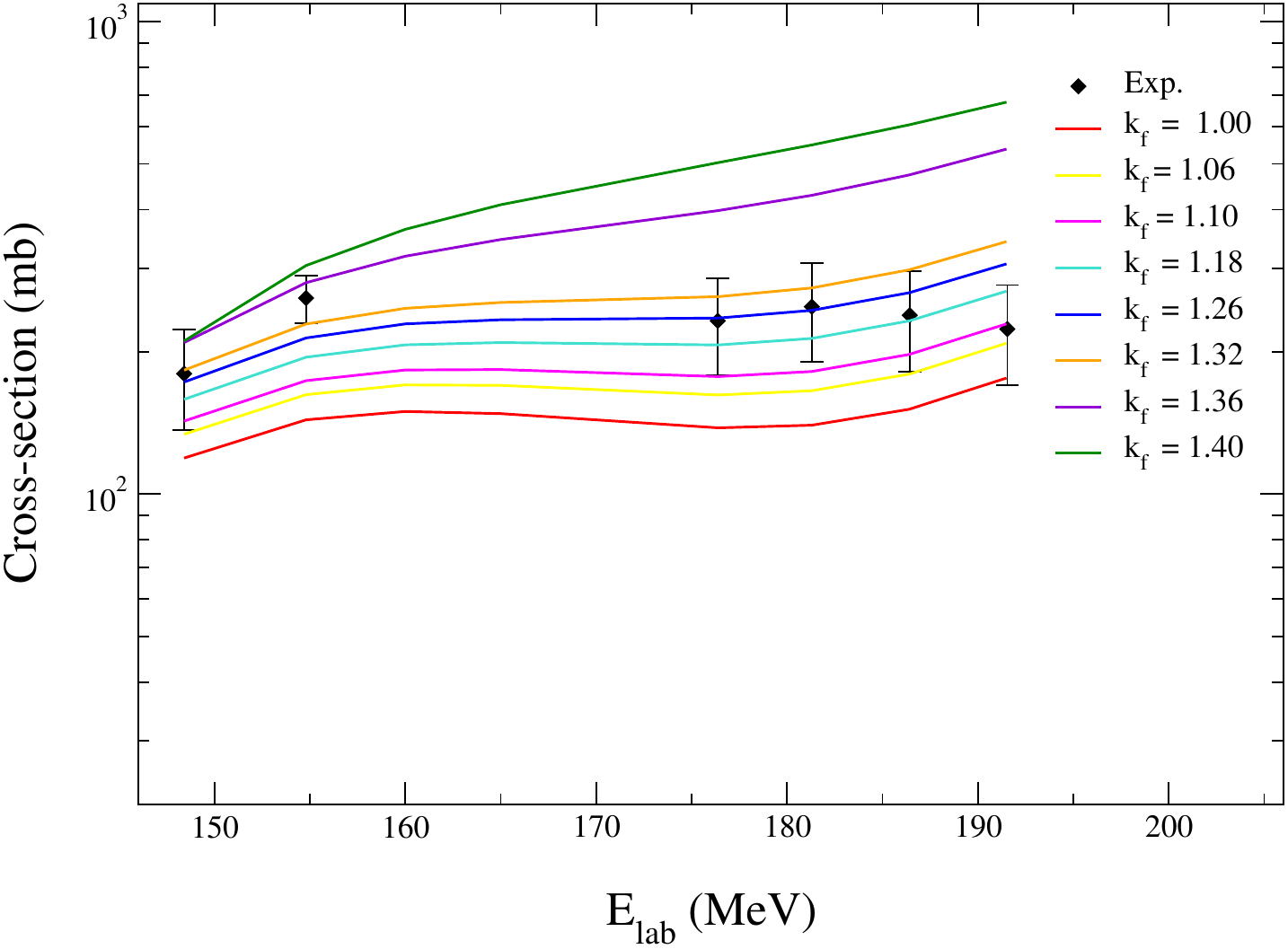}
\caption{(Color online)  Variation of the ER cross-sections for $^{32}$S +$^{154}$Sm with Sierk fission barrier (scaling factor $k_f$)  at different $E_{\rm lab}$.}
\label{fig_12}
\end{figure}
We now discuss the results of the CASCADE calculations to reproduce the ER cross-sections by varying the fission barrier and the viscosity parameter. The calculated ER cross-sections are shown and compared with the measured values in Figs. 12-14. While the present statistical model calculations can grossly reproduce the ER cross-sections, it is of primary importance to have some insight into the range of the vital parameters over which the data can be reproduced. In addition, for a highly dynamic process with multiple controlling factors, it is essential to have some idea about the relative contributions of the different parameters.  In the present analysis, we have tried to reproduce the ER data by varying both the fission barrier height as well as the viscosity parameter. The calculations were performed for both cases (i) with no viscosity ($\gamma=0$) and variable fission barriers and (ii) with variable viscosity parameter and a fixed Sierk barrier. As shown in Figs. \ref{fig_11} and \ref{fig_12}, the RLDM and the Sierk barrier were required to be scaled up by a factor $k_{f}$ to reproduce the ER cross-sections. In the case of the RLDM barrier, the experimental ER cross-sections were best matched with the calculations after up-scaling of the barrier by around 15\%, while in the case of the Sierk barrier, the scaling factor had to be changed by nearly 30\% to best fit the data. This is obvious from the fact that the RLDM is higher than the Sierk barrier up to 40 $\hbar$ angular momentum. The increase in the barrier height is expected to hinder the fission and lead to increased production cross-section of the ERs. The viscosity, inside the barrier, also plays the same role in inhibiting the flow of the flux towards the saddle and eventually increasing the ER cross-sections. We have performed the calculations by varying the Kramers' pre-saddle viscosity parameter $\gamma$ while keeping the fission barrier fixed. 
The outcome of these calculations is shown in Fig. \ref{fig_13}. In this calculation, a fixed Sierk barrier was considered while varying the dimensionless viscosity parameter $\gamma$ from zero to ten. While $\gamma$ = 0  is inadequate to reproduce the data, increasing the value up to 1 grossly reproduces the ER cross-sections at low energies. It is noteworthy that this low value of $\gamma$ is in conformity with previous analyses concluding much lower value of the viscosity parameter inside the saddle in comparison to the much larger values for the saddle-to-scission path \cite{2000_shaw}. We conclude this section by summarising the salient features of this phenomenological statistical model analysis. The basic motivation has been to do a realistic analysis and reduce the multi-dimensional parameters space, so as, not to stretch any of the parameters unrealistically. To this end, we started off by
feeding the experimentally determined spin distribution in the calculations. In order to have a realistic treatment of the Nuclear Level Density we have made use of the Ignatyuk-Reisdorf ansatz. We have carried out systematic analysis to estimate the possible range of the fission barrier and the viscosity parameter over which the ER data can be reproduced. After a comparative study of both the RLDM and Sierk barrier we zeroed upon the Sierk barrier as it is more advanced than the RLDM. Finally, fixing the Sierk barrier we can estimate the dimension-less viscosity parameter, required to produce the data, to be around 1.0.   
\begin{figure}[h]
\centering
\includegraphics[width=8.5cm]{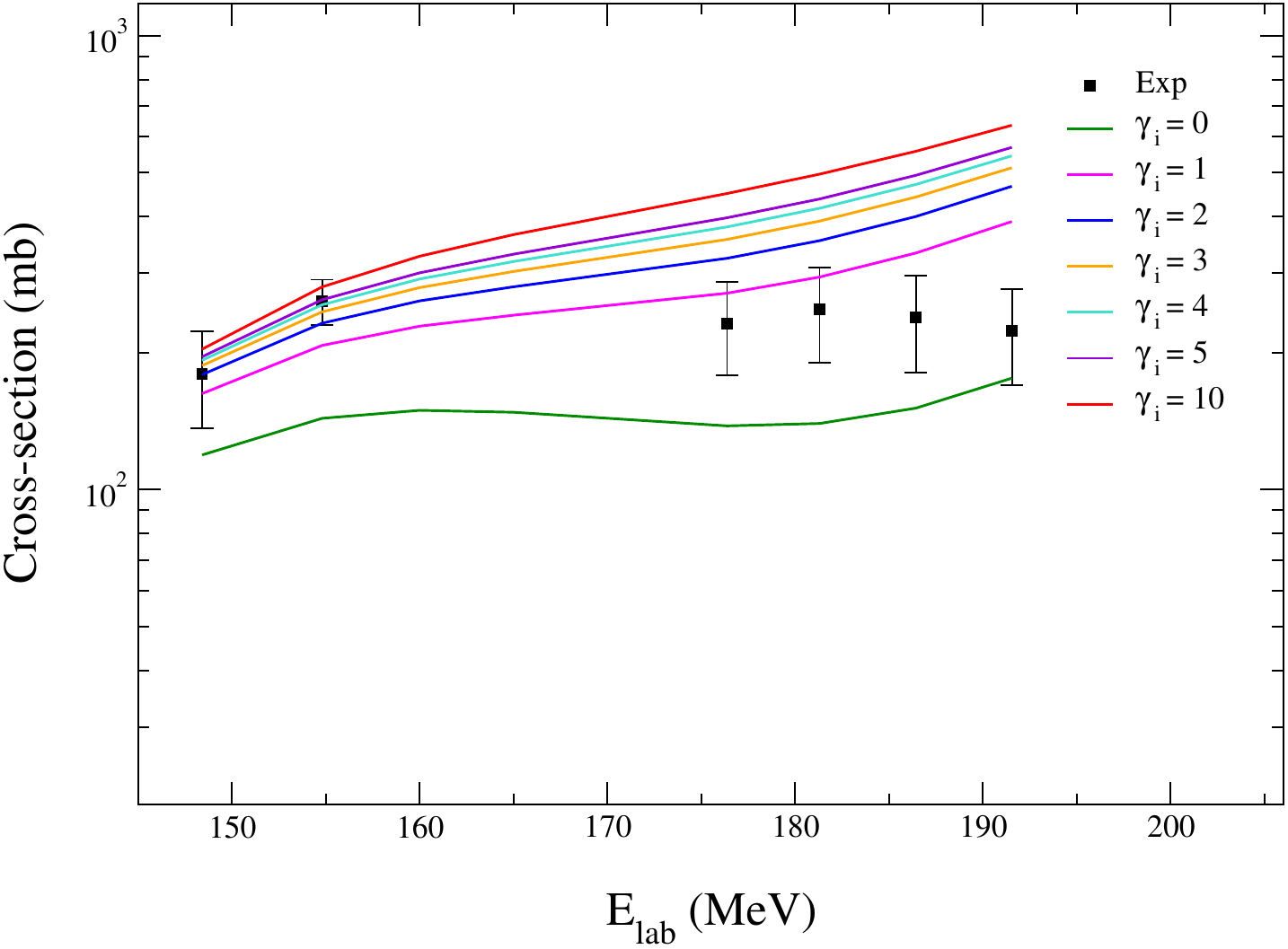}
\caption{(Color online)  Variation of the ER cross-sections for $^{32}$S +$^{154}$Sm with viscosity parameter $\gamma_i$ at different $E_{\rm lab}$.}
\label{fig_13}
\end{figure}

\subsection{\label{DNS_model}Dynamical model calculations}

In the DNS approach, the spin distribution of the CN formed in heavy ion collisions is calculated as the angular momentum distribution of the DNS formed at capture
of the projectile nucleus by the target nucleus \cite{Nasirov2005, Kayumov2022}. The cross-sections of the ERs formed after the emission of $x$-particles from the intermediate nucleus
with the excitation energy $E^*_{x}$ at each step $x$ of the de-excitation cascade is calculated as the sum of the partial cross-section of ER formation in collision with the orbital angular momentum $L=\ell\hbar $ by the formula
\cite{Mandaglio2012,Kayumov2022}:
\begin{eqnarray}\label{erpar}
    \sigma^{(x)}_{\rm ER}(E^*_{x})=\sum^{\ell_d}_{l=0}(2\ell+1)
    \sigma^{(x)}_{\rm ER}(E^*_{x},\ell),
\end{eqnarray}
where
\begin{eqnarray}\label{ercs}
    \sigma^{(x)}_{\rm ER}(E^*_{x},\ell)=\sigma^{x-1}_{\rm ER}(E^*_{x-1},\ell)W^x_{sur}(E^*_{x-1},\ell).
\end{eqnarray}
Here, $\sigma^{(x-1)}_{\rm ER}(E^*_{x-1},\ell)$ is the partial cross-section of the intermediate excited nucleus formation at the $(x-1)$th step and obviously,
\begin{equation}
\sigma^{(0)}_{\rm ER}(E^*_{\rm CN},\ell)=\sigma_{\rm fus}(E^*_{\rm CN},\ell),
\label{sigma0ER}
\end{equation}
$W^{(x)}_{sur}(E^*_{x-1},\ell)$ is the survival 	
probability of the $x$th intermediate nucleus against fission along the de-excitation cascade of CN. The survival probability $W^{(x)}_{sur}(E^*_{x-1},\ell)$ is
calculated by the statistical model contained in KEWPIE2 \cite{Kewpie2}, which is dedicated to the study of Super Heavy Elements (SHE).
 The CN excitation energy $E^*_{x}$  is calculated by taking into account the
 CN rotational energy $V^{\ell}_{\rm CN}$:
 \begin{equation}
E^*_{x}(\ell)=E_{\rm c.m.}+Q_{gg}-V^{\ell}_{\rm CN}-x\epsilon,
\end{equation}
where $Q_{gg}=B_P+B_T-B_{\rm CN}$ is a reaction energy balance; wherein 
$B_P$, $B_T$, and $B_{\rm CN}$ are the binding energies of the colliding nuclei and CN are obtained from the nuclear mass tables in Refs. \cite{Audi1995,Moller1995};
 $\epsilon$ is the emission energy of a particle;  $V^{\ell}_{\rm CN}$ is
 rotational energy of the compound nucleus and given by $V^{\ell}_{\rm CN}=5 L(L+1)/(2m A_{\rm CN}(a_{\rm CN}^2+b_{\rm CN}^2))$, where $m$ is a nucleon mass; 
$a_{\rm CN}$ and $b_{\rm CN}$ are the long and short axes of the ellipsoid compound 
nucleus. The partial fusion excitation function used in Eq. (\ref{sigma0ER}) is  determined by the partial capture cross-sections $\sigma^{\ell}_{cap}(\alpha_i,\beta_i)$ and   fusion  probabilities
$P_{CN}(E_{\rm c.m.},\ell,\{\alpha_i,\beta_i\})$ of DNS for the heavy collisions
with the given orbital angular momentum $\ell$:
\begin{align}
\label{parfus}
\sigma_{\rm fus}(E^*_{\rm c.m.},\ell,\{\alpha_i,\beta_i\})=
\sigma_{cap}(E^*_{\rm c.m.},\ell;\{\alpha_i,\beta_i\}) \nonumber \\
P_{\rm CN}(E^*_{\rm c.m.},\ell;\{\alpha_i,\beta_i\}).
\end{align}
In case of collision of the spherical and deformed nuclei, for example, for the $^{32}$S + $^{154}$Sm reaction, a partial fusion cross-section is found by averaging over all values of the orientation angles $\alpha_2$ symmetry axis of
the deformed $^{154}$Sm  nucleus:
\begin{align}\label{parfusal}
\sigma_{fus}(E^*_{\rm c.m.},\ell,\alpha_1,\beta_1)&=&\int\limits_{0}^{\pi/2}
\sigma_{fus}(E^*_{\rm c.m.},\ell;\alpha_1,\alpha_2)\nonumber\\
&&\times\sin{\alpha_2}d\alpha_2,
\end{align}
and by averaging over the surface vibrational states $\beta_1$ of the spherical
$^{32}$S as independent harmonic vibrations and the nuclear radius is taken to be distributed as a Gaussian distribution~\cite{EsbensenNPA352},
\begin{equation}
g(\alpha_1,\beta) = {\rm exp}
\left[-\frac{R_0^2(\sum_{\lambda}\beta_{\lambda} Y_{\lambda 0}^* (\alpha_1))^2}{2 \sigma_{\beta}^2} \right] (2\pi \sigma_{\beta}^2)^{-1/2},
\end{equation}
where $\alpha_1$ is the direction angle of the axis along which the surface of
the spherical nucleus vibrates and deformation parameters are changed between
  the negative and positive values around the spherical shape. For simplicity, we use $\alpha_1=0$:
\begin{equation}
\sigma^2_{\beta_1} = R_0^2 \sum_{\lambda}\frac{2\lambda + 1}{4\pi} \frac{\hbar}{2D_\lambda \omega_\lambda} = \frac{R_0^2}{4\pi} \sum_{\lambda} \beta_\lambda^2,
\end{equation}
where $\omega_{\lambda}$ is the frequency and $D_{\lambda}$ is the mass parameter of a collective mode.

As the amplitudes of the surface vibration, we use the deformation parameters of the first excited 2$^+$ and $3^-$ states of the colliding nuclei. The values of the deformation parameters ($\beta^+_2$) and ($\beta^-_3$) of the first excited 2$^+$ and $3^-$ are presented in Table \ref{tabdeform} which are taken from Ref(s).~\cite{Raman} and \cite{Spear}, respectively.
\begin{table}[b]
	\caption{Deformation parameters $\beta_2$ and $\beta_3$ of first excited
	2$^+$ and $3^-$ states of the colliding nuclei $^{32}$S and $^{154}$Sm
 used in the calculations in this work.\\
\centering
		\label{tabdeform}}
\begin{tabular}{cccc}
\hline
\hline
				Nucleus  &  $^{32}$S &  $^{154}$Sm  \\ [1ex]
				\hline
				$\beta^+_2$~\cite{Raman} & 0.312  & 0.34   \\ [1ex]
				$\beta^-_3$~\cite{Spear}  & 0.410    & 0.13   \\ [1ex]

\hline
\hline
\end{tabular}
\end{table}

Partial fusion cross-section is found by averaging over values of the vibrational states $\beta_i$  of the spherical nuclei ($\alpha_1=0$):
\begin{equation}
\label{parfusvib}
\sigma_{\rm fus}(E^*_{\rm c.m.},\ell)=\int\limits_{-\beta_0^{(1)}}^{+\beta_0^{(1)}}g_1(\beta_1)d\beta_1
\sigma_{\rm fus}(E^*_{\rm c.m.},\ell;\beta_1).
\end{equation}

The cross-sections of the ER formation in
  $x$n channels of the incomplete fusion (ICF) \cite{nasirov_2022, nasirov_2023} accompanied by the emission
  of the $\alpha$ particle have been calculated by replacing
  the fusion probability $P_{\rm CN}$
  of nuclei  with the probability of the $\alpha$-particle emission
  in Eq. (\ref{parfus}):
  \begin{align}
\label{paricfus}
\sigma_{\rm ICF}(E,\ell,\{\alpha_i,\beta_i\})=
\sigma_{cap}(E^*_{\rm c.m.},\ell;\{\alpha_i,\beta_i\})\nonumber \\
Y_{\rm Z}(E^*_{\rm c.m.},\ell;\{\alpha_i,\beta_i\}).
  \end{align}

In the case of the ICF, the excitation energy $E^*_{\rm ICF}$ of the system at the break
up is calculated using the expression:
\begin{align}
E^*_{Z}(E_{\rm c.m.},A,L,\{\beta_i,\alpha_i\})=E_{\rm c.m.}+\Delta Q_{\rm gg}(Z,A) \nonumber \\
-V(Z,A,R_m,L,\{\beta_i,\alpha_i\}),
\label{Edns}
\end{align}
where $\Delta Q_{\rm gg}(Z,A)=B+B_c-B_P-B_T$;
$B$ and $B_c$  are binding energies of the $\alpha$ particle ($Z=2$ and $A=4$) and
conjugate nucleus, respectively; $V(Z,A,R_m,L,\{\beta_i,\alpha_i\})$  is the minimum value of the potential well of the $\alpha$-nucleus interaction.  Details of calculation of $V(Z,A,R,L,\{\beta_i,\alpha_i\})$ are given in Refs. \cite{Nasirov2005,Kayumov2022}.
\begin{figure}[h]
\centering
\includegraphics[width=9.2cm]{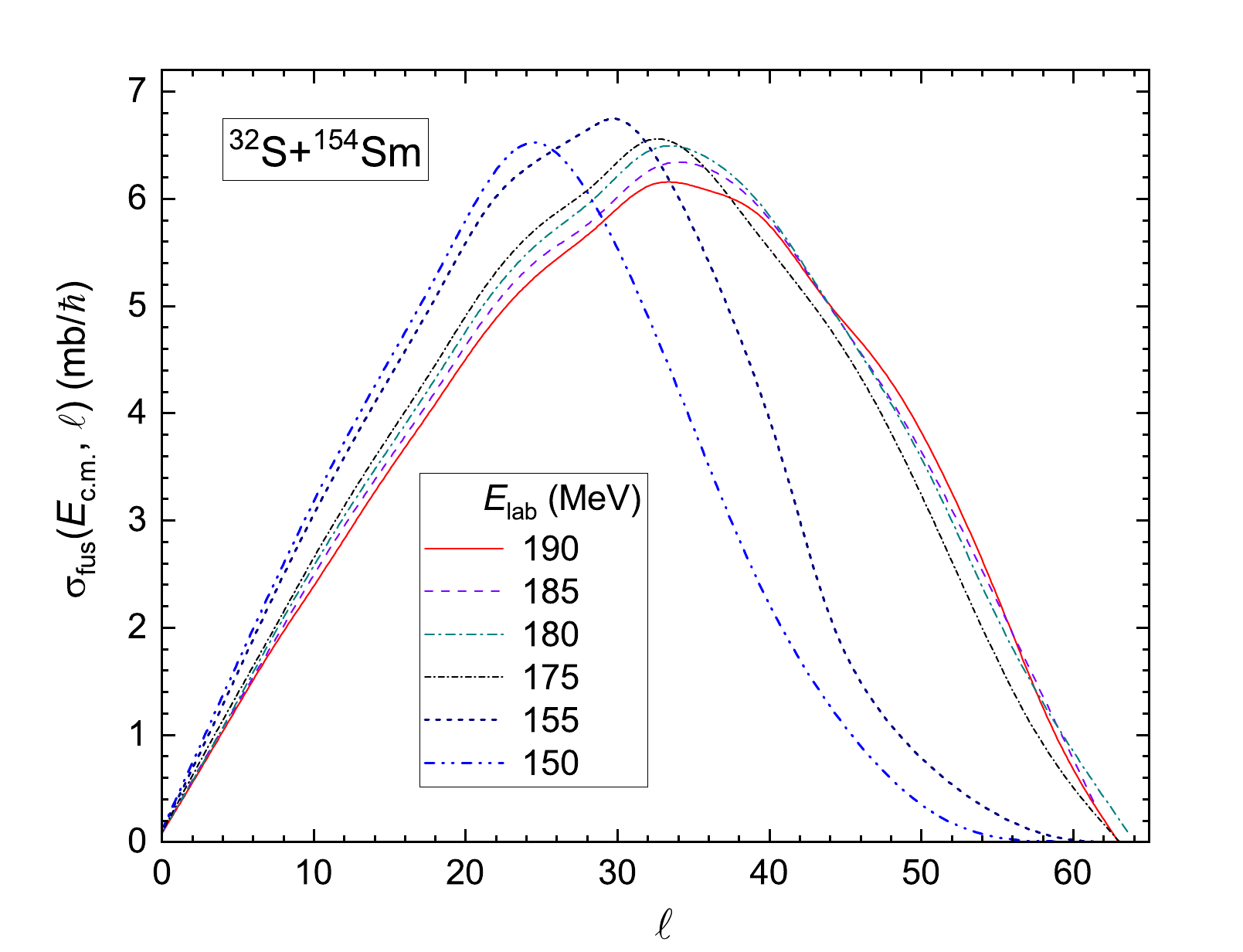}
\caption{(Color online) The  partial fusion cross-section
$\sigma_{\rm fus}(E_{\rm c.m.},\ell)$ calculated in this work for the
$^{32}$S+$^{154}$Sm reaction as a function of the collision energy
$E_{\rm c.m.}$ and orbital angular momentum $\ell$.}
\label{fig_14}
\end{figure}
Fig. \ref{fig_14} shows the partial fusion cross-section $\sigma_{\rm fus}$ calculated
 by Eq. (\ref{parfusvib}). Its values have been used in calculating the ER cross 
sections for the $x$n emission channels  by Eq. (\ref{erpar}). The spin distribution of 
the compound nucleus formed at the complete fusion depends on the beam energy 
$E_{\rm c.m.}$. But the values of $\ell$ in the range $20<\ell<50 $ give the main 
contribution to the complete fusion. The qualitative agreement between theoretical
results of $\sigma_{\rm fus}(E_{\rm c.m.}, \ell)$ (see Fig. \ref{fig_14}) and spin distribution
obtained in the experimental data presented in Fig. \ref{fig_9} are fairly well for the two lowest beam energies ( 150 and 155 MeV ). 

The large values of $\ell$ of the compound nucleus are restricted by the strong decrease of the fission barrier $B_f$ as a function
of $\ell$ according to Fig. \ref{fig_10}.
The ER cross-section calculated as a sum of the ones found for the $x$n channels Eq. (\ref{erpar}) at de-excitation of  the heated and rotating compound nucleus formed
in complete fusion:
\begin{equation}
\sigma_{\rm ER}(E_{\rm c.m.})=\sum_x \sigma^{(x)}_{\rm ER}(E^*_{x}).
\end{equation}
 
\begin{figure}[h]
\centering
\includegraphics[width=9.2cm]{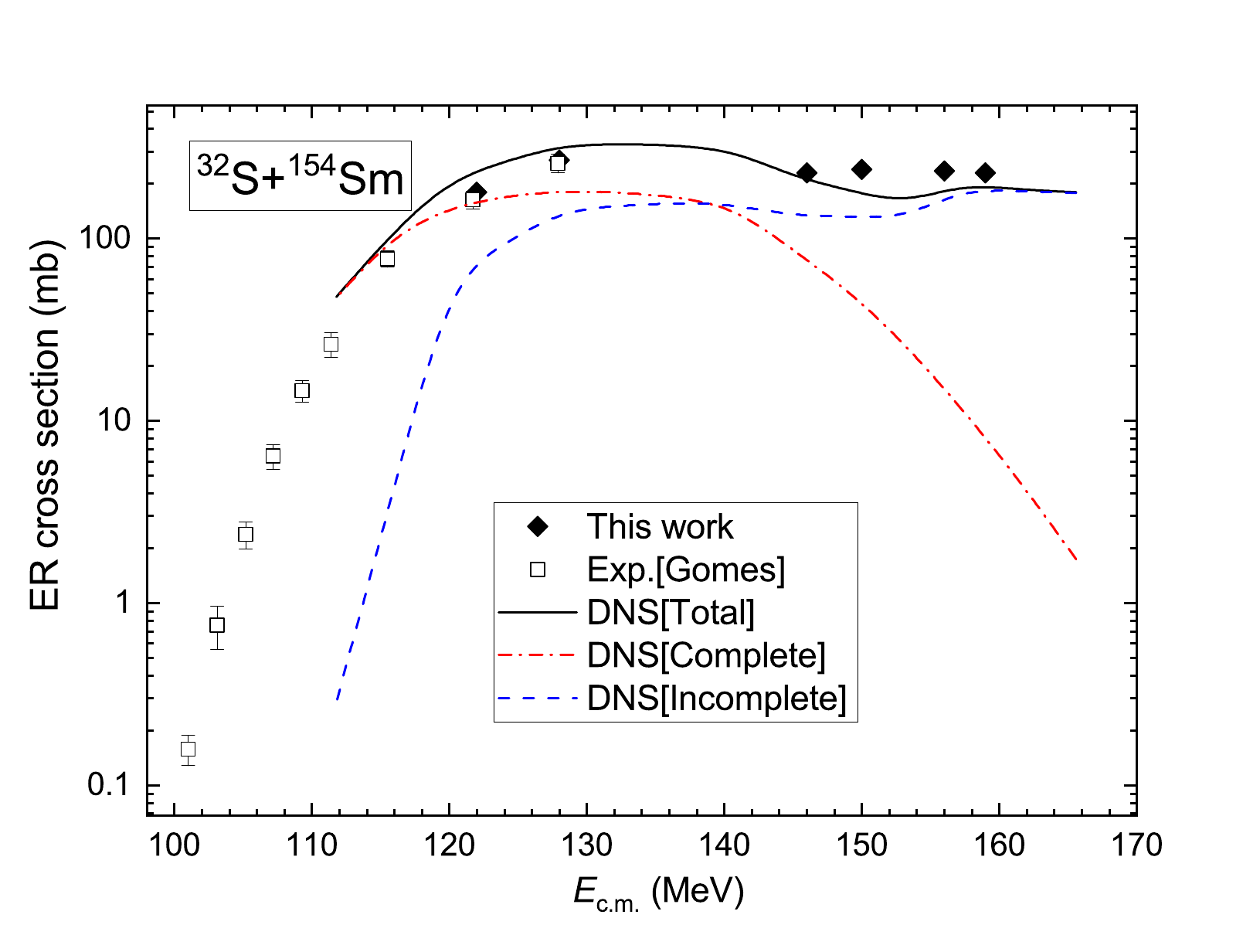}
\caption{(Color online) Comparison of the theoretical  (solid curve)
and measured the total cross-sections of the evaporation residues formed in
the $^{32}$S+$^{154}$Sm reactions. The filled diamonds and open squares  present
the experimental data of this work and the ones obtained from Ref. \cite{gomes_1994},
respectively. The dashed and dot-dashed curves are contributions of the
complete and incomplete fusion mechanisms estimated by the DNS model,
respectively.
 }
\label{fig_16}
\end{figure}

 The results of the calculation are presented in Fig. \ref{fig_16}.
 The curve of $\sigma_{\rm ER}(E_{\rm c.m.})$ corresponding to the 
 complete fusion decreases at $E_{\rm c.m.} $ above 140 MeV while the experimental values of $\sigma_{\rm ER}$ do not decrease.  
 This behavior is related to decrease
of the fission barrier as a function of the excitation energy $E^*_{\rm CN}$ of
the compound nucleus. 
\begin{figure}[h]
\centering
\includegraphics[width=9.2cm]{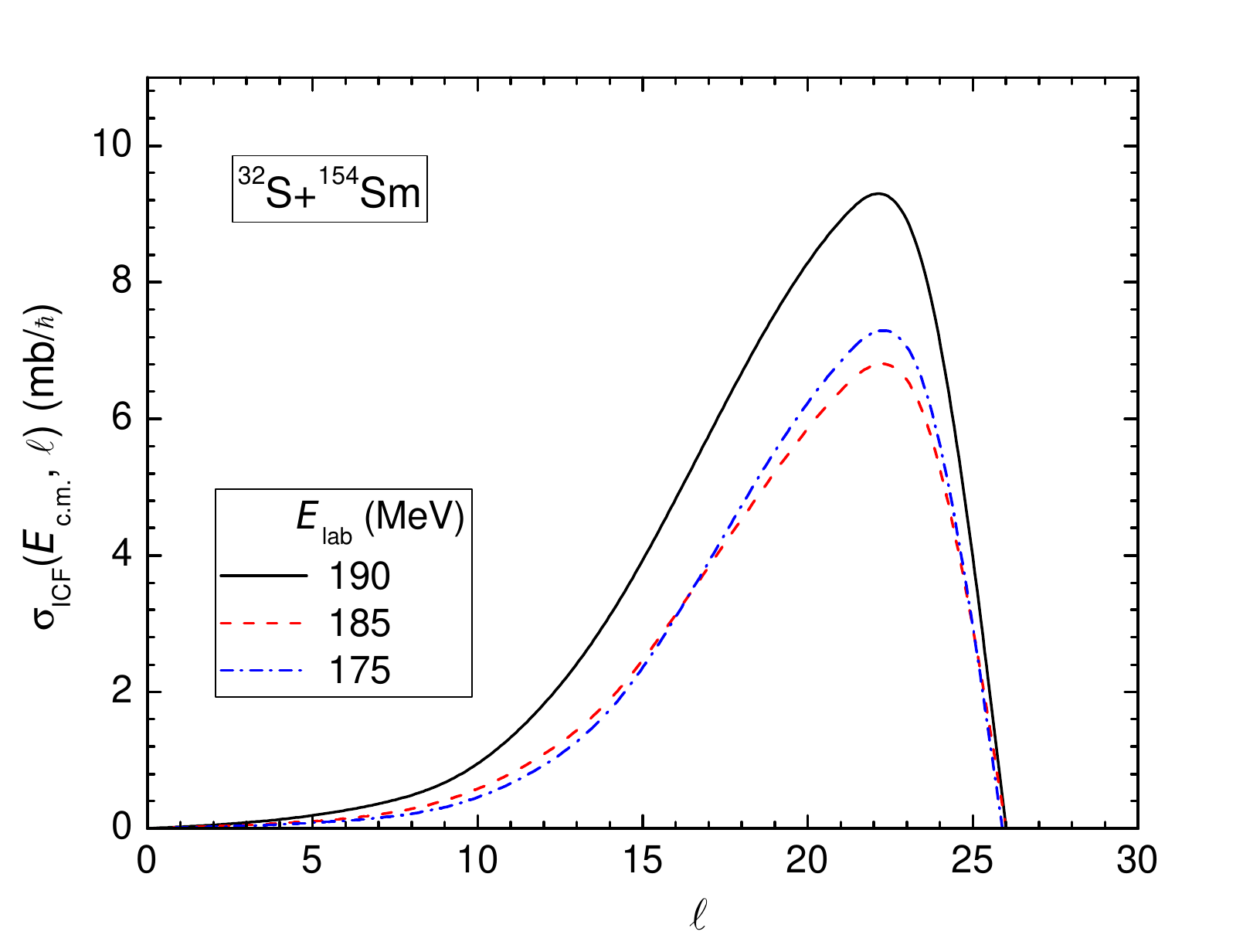}
\caption{(Color online) The partial incomplete fusion cross-section
$\sigma_{\rm ICF}(E_{\rm c.m.},\ell)$ calculated in this work for the
$^{32}$S+$^{154}$Sm reaction as a function of the collision energy
$E_{\rm c.m.}$ and orbital angular momentum $\ell$.}
\label{fig_15}
\end{figure}

 To solve this problem, we have performed calculations to estimate  the contribution 
 of the ER cross-section related to the incomplete fusion mechanism.
 For this, we have used an incomplete fusion cross-section Eq.
  (\ref{paricfus}) instead of the complete fusion cross-section in Eq. (\ref{sigma0ER}). The partial cross-sections of the incomplete fusion show the ``$\ell$-window'' 
 $20 < \ell < 40$ for the values of the angular momentum which lead to incomplete fusion 
 (Fig. \ref{fig_15}). 
  The calculated results of $\sigma^{\rm ICF}_{\rm ER}(E_{\rm c.m.})$  are presented 
  in Fig. \ref{fig_16} by the dashed blue line. 
  Its contribution is dominant at larger values of $E_{\rm c.m.}$ and 
  this behavior is inherent for the incomplete fusion reactions \cite{Agarwal}. 

\section{\label{sec:level5}Discussion} The discussion in this section is centered around the salient features emerging out of the present measurements and analysis. This is the second reported measurement for ER cross-sections from the $^{186}$Pt compound nucleus formed by $^{32}$S+$^{154}$Sm reaction. The previous measurements by Gomes \textit{et al.}  \cite{gomes_1994} have reported ER cross-sections just about 10$\%$ above the barrier. In the present work, five new data points going up to around 100 MeV excitation energy have been added. In addition, we have also, for the first time, measured the $\gamma$-fold distributions and extracted the spin distributions for this system. The statistical model calculations have been carried out  to reproduce the ER cross-sections. The two primary physical quantities controlling the production cross-sections of the ER are barrier height and nuclear viscosity. An attempt has been made to fit the experimental cross-sections by varying both these quantities and get some idea about the realistic range of the parameter spaces. The reproduction of ER cross-sections is achieved by scaling the fission barrier ($K_f$) within the range of 1.00-1.21 for the RLDM barrier and 1.00-1.40 for the Sierk barrier. The best fit scaling factors are around 1.15 and 1.30 for the RLDM and Sierk barriers, respectively.  We have chosen to use the Sierk barrier for the calculations. Keeping the Sierk barrier fixed, the viscosity parameter ($\gamma$) is varied over a range of 0 to 10. The data is reasonably well reproduced for $\gamma$ around 1.0.  In the present phenomenological calculations, the viscosity parameter controlling the ER cross-section is the Kramers' $\gamma$ value inside the barrier. The statistical model calculations have also been complemented by dynamical calculations within the framework of the DNS model. This model calculation shows the role of both complete and incomplete fusion channels in reproducing the total ER cross-sections. It is shown that the ER cross-sections from the complete fusion channel go down beyond $E_{\rm c.m.}$ = 140 MeV. In contrast, the cross-sections due to an incomplete fusion channel do not go down with beam energy. The cross-sections are well reproduced by considering both complete and incomplete fusion processes.   

\begin{figure}[h!]
 \centering
  \includegraphics[width=8.5cm]{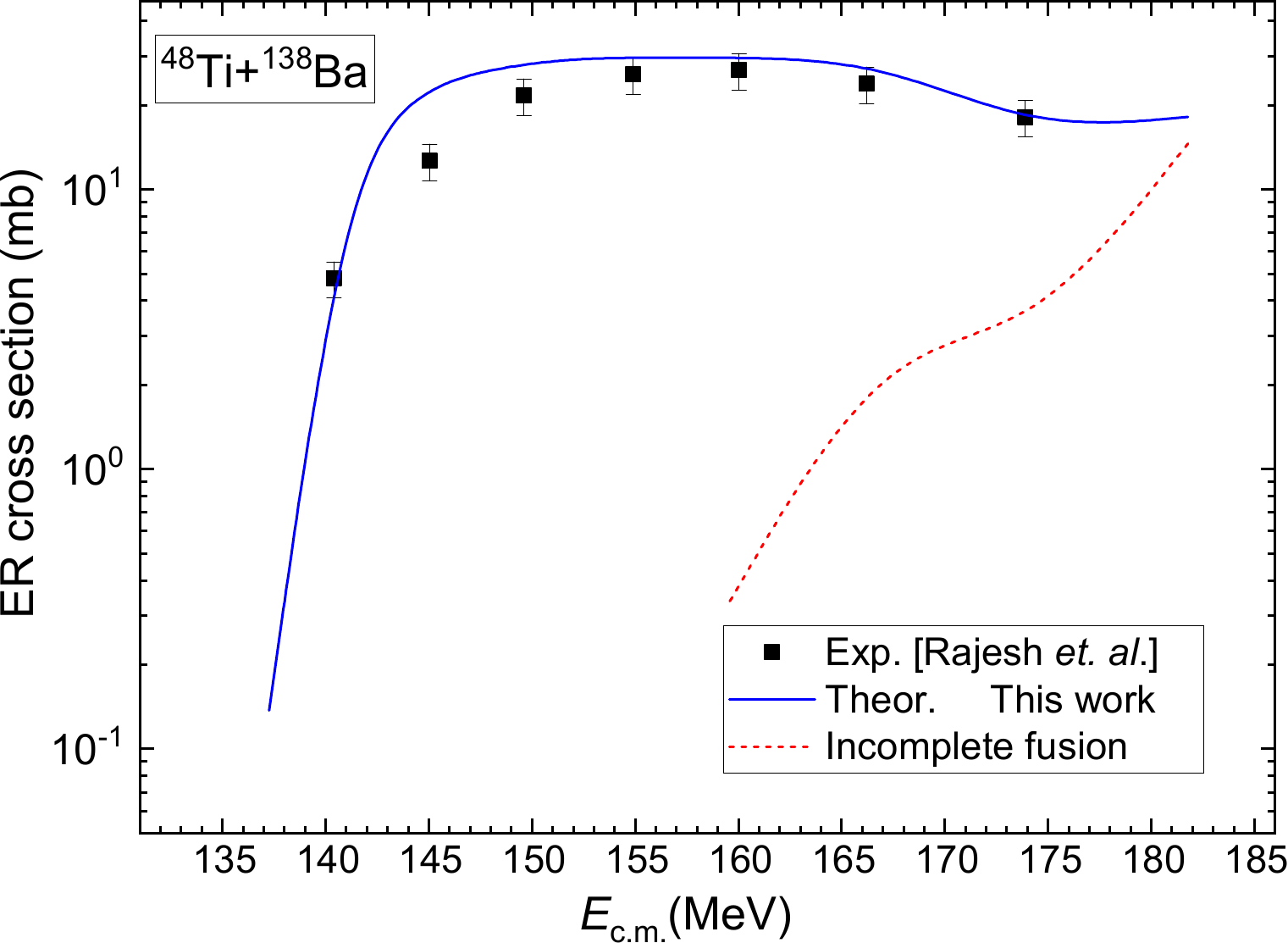} 
    \caption{  (Color online) Comparison of the theoretical  (solid curve)
and measured the total cross-sections of the evaporation residues formed in
the $^{48}$Ti + $^{138}$Ba reactions. The dashed curves are contributions of the incomplete fusion mechanisms estimated by the DNS model.}
   \label{fig_18}
\end{figure}

\section{\label{sec:level6}Summary}
Compound nucleus $^{186}$Pt has been populated using $^{32}$S +$^{154}$Sm reaction. The ER cross-sections and ER-gated spin distributions have been measured for the first time at six different beam energies above the barrier using the HYRA gas-filled recoil separator and the TIFR 4$\pi$ sum-spin spectrometer.  Theoretical calculations have been done using both statistical and dynamical model calculations. The statistical model calculations have generated the range of parameter space for both the fission barrier height and the nuclear viscosity parameter over which the ER cross-section data can be reproduced.  The DNS model calculations reproduce the data considering both complete and incomplete fusion processes. An important observation of the work is the clear difference in the ER cross-sections of the $^{186}$Pt compound nucleus compared to previous measurements for the same compound nucleus populated by a very different target-projectile combination with much less mass asymmetry. This further demonstrates the role of quasi-fission hindering complete fusion and less production of ER compared to an entrance channel with larger mass asymmetry.
 \section*{Acknowledgments}
One of the authors (R. Sariyal) acknowledges the Department of Science \& Technology (DST), Govt. Of India for the INSPIRE fellowship. We thank the Pelletron and LINAC group of IUAC for their support during the entire run of the experiment and for delivering a high-quality beam. We also acknowledge the support of the target laboratory of IUAC.

\end{document}